\documentclass[useAMS,usenatbib]{mn2e}

\usepackage{xcolor}
\usepackage{graphicx}
\usepackage{hyperref}
\usepackage{amsmath}
\usepackage{eulervm}
\usepackage{ulem}
\usepackage{color}

\title[\mbox{Mg{\sc\ II}} absorbers in SDSS DR12Q spectra] {Intervening \mbox{Mg{\sc\ II}} absorption systems from the SDSS DR12 quasar spectra}
\author[Srinivasan Raghunathan et al.]{Srinivasan Raghunathan$^{1,2}$\thanks{send correspondence to Srinivasan R., sri@das.uchile.cl}, Roger G. Clowes$^{3}$, Luis E. Campusano$^1$, \and Ilona K. S{\"o}chting$^4$, Matthew J. Graham$^{5,6}$, and Gerard M. Williger$^{3,7,8}$\\\\
$^{1}$Departamento de Astronom\'{i}a, Universidad de Chile, Camino del Observatorio 1515, Santiago, Chile\\
$^{2}$School of Physics, University of Melbourne, Parkville VIC 3010, Australia\\
$^{3}$Jeremiah Horrocks Institute, University of Central Lancashire, Preston PR1 2HE, UK\\
$^{4}$Astrophysics, Denys Wilkinson Building, Keble Road, University of Oxford, Oxford OX1 3RH, UK\\
$^{5}$California Institute of Technology, 1200 East California Boulevard, Pasadena, CA 91125, USA\\
$^{6}$National Optical Astronomy Observatory, 950 N Cherry Avenue, Tucson, AZ 85719\\
$^{7}$Department of Physics \& Astronomy, University of Louisville, KY 40292, USA\\
$^{8}$Institute for Astrophysics and Computational Sciences, The Catholic University of America, DC 20064, USA}
\begin{document}

\date{Accepted 2016 August 16. Received 2016 August 8; in original form 2015 September 8}

\pagerange{1--15} \pubyear{2016}

\maketitle

\label{firstpage}

\begin{abstract}
We present the catalogue of the \mbox{Mg{\sc\ II}} absorption systems detected at a high significance level using an automated search algorithm in the spectra of quasars from the twelfth data release of the Sloan Digital Sky Survey. A total of 266,433 background quasars were searched for the presence of absorption systems in their spectra. The continuum modelling for the quasar spectra was performed using a mean filter. A pseudo-continuum derived using a median filter was used to trace the emission lines. The absorption system catalogue contains 39,694 \mbox{Mg{\sc\ II}} systems detected at a \mbox{6.0, 3.0$\sigma$} level respectively for the two lines of the doublet. The catalogue was constrained to an absorption line redshift of \mbox{0.35 $\le$ z$_{2796}$ $\le$ 2.3}. The rest-frame equivalent width of the $\lambda$2796 line ranges between \mbox{0.2 $\le$ W$_r$ $\le$ 6.2 \AA}. Using Gaussian-noise only simulations we estimate a false positive rate of 7.7 per cent in the catalogue. We measured the number density $\partial N^{2796}/\partial z$ of \mbox{Mg{\sc\ II}} absorbers and find evidence for steeper evolution of the systems with \mbox{W$_r \ge$ 1.2 \AA} at low redshifts (z$_{2796}$ $\le$ 1.0), consistent with other earlier studies. A suite of null tests over the redshift range \mbox{0.5 $\le$ z$_{2796}$ $\le$ 1.5} was used to study the presence of systematics and selection effects like the dependence of the number density evolution of the absorption systems on the properties of the background quasar spectra. The null tests do not indicate the presence of any selection effects in the absorption catalogue if the quasars with spectral signal-to-noise level less than 5.0 are removed. The resultant catalogue contains 36,981 absorption systems. The \mbox{Mg{\sc\ II}} absorption catalogue is publicly available and can be downloaded from the link \url{http://srini.ph.unimelb.edu.au/mgii.php}.
\end{abstract}


\section{Introduction}
\label{introduction}
Quasars are extremely luminous light sources and can be used to study the high redshift universe as they can be easily observed using ground-based telescopes. The observed light from distant quasars is altered when crossing cold gas belonging to objects on its way to the Earth and can therefore reveal information about both the gas presence and the properties. Distinct absorption patterns in the quasar spectra can be produced due to photon interaction with the intervening gas, either neutral or ionised, providing the basis for Quasar Absorption Line studies (QALs). QALs contribute both to the characterisation of faint objects in the quasar line-of-sight (LOS) including some that cannot be detected directly with the telescopes, and the acquisition of unbiased one-dimensional information of the highly ionised intergalactic (IGM) and the intra-cluster medium(ICM). Since 90 per cent of the baryons in the universe are in the form of non-luminous gas, QALs are one of the few cosmological observations to trace the baryons at high redshifts. Other indirect measurements include the Cosmic Microwave Background Radiation \citep{penzias1965,hinshaw2013,planck2015}, the weak gravitational lensing \citep{bartelmann1999, zaldarriaga2000}, the Baryonic Acoustic Oscillations \citep{bassett2009}, and the 21-cm hydrogen line measurements \citep{pritchard2012}.

There are several classes of QALs and a typical quasar spectrum can contain hydrogen (HI) or metal lines because of the chance alignment of several astrophysical sources along the quasar LOS. While metal lines like \mbox{C{\sc\ IV}}, FeII, \mbox{Mg{\sc\ II}}, etc. trace galaxies, the HI lines can represent either a galaxy or the IGM depending on their column densities \citep{narayanan2008}.

The \mbox{Mg{\sc\ II}} doublets can be used to study the gaseous components of galaxies as they trace the low-ionisation gas with column densities \mbox{10$^{16} <$ N(H{I}) $\le$ 10$^{22}$ cm$^{-2}$} \citep{kacprzak2011}. The \mbox{Mg{\sc\ II}} doublets are classified into strong (W$_r$ $\ge$ 0.3 \AA) and weak (W$_r < $ 0.3 \AA) systems based on their rest-frame\footnote{Throughout this paper the subscript \textit{r} will represent the rest-frame value which is the observed value divided by (1+z).} equivalent widths (EWs). The ease of detection of the \mbox{Mg{\sc\ II}} doublets, the ability of strong systems to trace neutral hydrogen gas, and their association with galaxies and star formation history \citep{nestor2005, mshar2007, narayanan2008, tinker2010} make them ideal candidates for QAL studies compared to other metal lines. The \mbox{Mg{\sc\ II}} doublets have rest-frame wavelengths ($\lambda_r$) of 2796, 2803 \AA\ and hence can be easily observed using ground-based telescopes from redshifts as low as z=0.11. In the optical spectra like the Sloan Digital Sky Survey (SDSS), the doublet can be observed in the redshift range $0.35 \le z \le 2.3 $.

In this paper, we describe the catalogue of 39,694 \mbox{Mg{\sc\ II}} doublets detected at a 6$\sigma$ level in the spectra of SDSS DR12 quasars (DR12Q) using an automated search algorithm. Similar \mbox{Mg{\sc\ II}} absorption system searches have been done by several groups using the earlier Data Releases (DR) of the SDSS quasar spectra \citep{bouche2004, nestor2005, prochter2006, narayanan2008, lundgren2009, quider2011, seyffert2013, zhu2013}. There were also \mbox{Mg{\sc\ II}} studies from other surveys of quasars \citep{lanzetta1987, charlton1998, churchill1999, ellison2004} and Gamma-Ray Bursts \citep{prochter2006b, tejos2009}. The absorption line detection method in the above surveys was either visual or automatic with various levels of visual checks depending on the quality and the number of quasar spectra used. The catalogue presented in this work along with \citet{zhu2013} (ZM13 hereafter) are the only fully automatic absorption system catalogues using the SDSS quasar spectra. 

The increasing number of quasar samples with every new data release of the SDSS, because of better statistics, leads to a better determination of the cosmological evolution of the \mbox{Mg{\sc\ II}} absorbers as well as to the identification of more homogenous samples of \mbox{Mg{\sc\ II}} absorbers for use as probes to study the galaxies and the Large-Scale Structures (LSS) of the universe. The \mbox{Mg{\sc\ II}} systems are important in understanding galaxy evolution and LSS. For example: \citet{lundgren2009} and \citet{gauthier2009} have employed the \mbox{Mg{\sc\ II}} absorbers to study the clustering of Luminous Red Galaxies; \citet{nestor2011} have used them to identify outflows from high redshift starburst galaxies; \cite{lopez2008} used them to investigate the galaxy cluster environment; \citet{williger2002} and \cite{clowes2013} provided independent corroboration of even larger structures called the Large Quasar Groups using the \mbox{Mg{\sc\ II}} absorption systems. 

The \mbox{Mg{\sc\ II}} catalogue described in this paper is a publicly available general purpose catalogue. The paper is organised as follows. The quasar spectra sample, continuum and noise estimation, and the automatic search algorithm are explained in \S2. The catalogue refinement, cuts applied to eliminate spurious detections, estimation of false positives, and the caveats are explained in \S3. The results of the survey -- catalogue description, statistical properties and the cosmological evolution of the \mbox{Mg{\sc\ II}} systems, and null tests for systematics study -- are in \S4, and \S5. We conclude in \S6.

\section{Method}
\subsection{SDSS DR12Q sample}
The DR12 is the final data release of the SDSS-III \citep{eisenstein2011,dawson2013} covering $\sim$9376 deg$^2$ of the sky in total \citep{alam2015}. The survey was carried out over a period of 14 years using a 2.5m dedicated optical telescope situated at the Apache Point Observatory in New Mexico \citep{gunn2006}. The DR12Q catalogue\footnote{\url{http://www.sdss.org/dr12/algorithms/boss-dr12-quasar-catalog/}} \citep{paris2015} contains 297,301 quasar spectra in total and 295,944 of them are at high redshift \mbox{(0.35 $\le$ z $\le$ 7.0; $\langle z_{QSO} \rangle \sim 2.15$)} desirable for the \mbox{Mg{\sc\ II}} doublet detection. The 29,580 quasars flagged as broad absorption line (BAL) quasars by the SDSS were not considered for the \mbox{Mg{\sc\ II}} search in this work. The final list in which absorption systems were searched consisted of 266,433 quasars. The spectra cover almost the entire optical window \mbox{(3500 \AA\ - 10500 \AA)} with a resolution \mbox{$\lambda/\Delta\lambda$} ranging from 1500 to 2500 \citep{smee2013}. The spectra were downloaded directly from the SDSS webpage\footnote{\url{http://data.sdss3.org/sas/dr12/boss/spectro/redux/}}. 

\subsection{Continuum estimation}
\label{sec_continuum_noise}
\begin{figure}
\centering
\includegraphics[width=0.5\textwidth, height=0.36\textwidth,clip=]{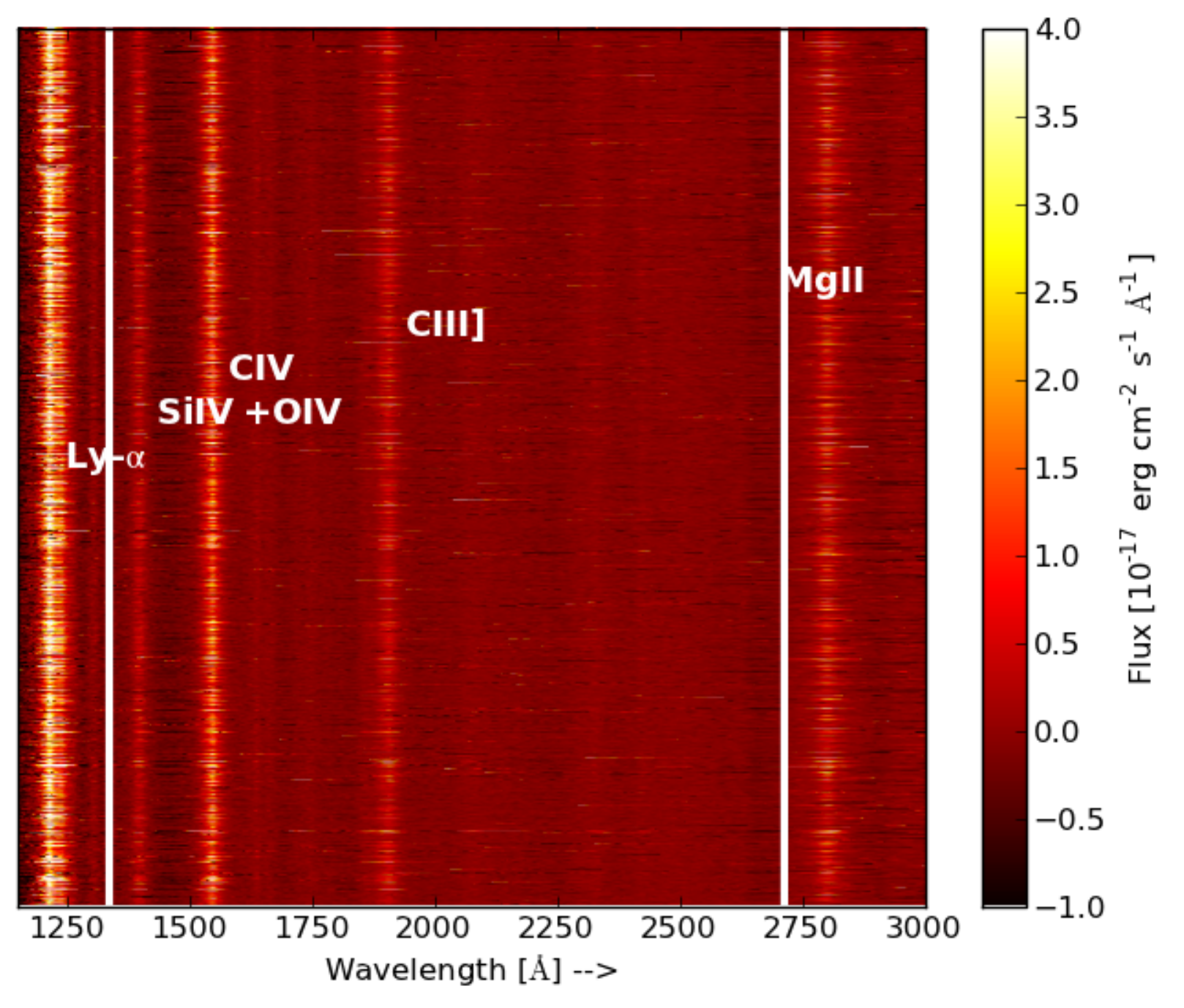}
\caption{A waterfall plot showing the difference between the final continuum, and the TM filter continuum for 10000 randomly selected non-BAL DR12Q spectra. The bright patches near the marked emission lines indicate that they are traced by the pseudo-continuum. The white vertical lines mark the \mbox{Mg{\sc\ II}} absorption search window (see text for more details) in the current work.}
\label{fig_con_diff}
\end{figure}

A basic prerequisite to perform the spectral analysis is to determine a satisfactory continuum. The continuum fit in this work was computed through a mean filter algorithm which utilises wavelength windows of varying size (see below) along the spectrum. In the regions of an SDSS quasar spectrum without emission lines, the continuum derived using the mean filter is the adopted one. For the quasar spectral regions containing strong emission lines (see Fig. \ref{fig_con_diff}) a special procedure has to be adopted because some intervening absorption lines can actually be on top of these emission lines. In order to detect absorption systems in the emission line regions, a pseudo-continuum that follows the emission lines was determined using a median filter algorithm.

The mean filter used is the modified Thompson-Martin (TM) digital filter \citep{thompson1971,martin1979} as modified by \cite{clowes1983}. The TM filter works by averaging flux values over several pixels (specified by a window size) based on the following procedure.
\begin{description}
\item[1.] Create a window with width $W_{width}$ centered on each pixel of the spectrum containing $N_{pix}$ pixels and record the window $W_{max}$ containing the maximum deviation.
\item[2.] Average the flux values in $W_{max}$.
\item[3.] Steps 1 and 2 are repeated $W_{pass}$ times until the entire spectrum is averaged out for the chosen window size.
\item[4.] Repeat steps 1 through 3 for $W_{total}$ window sizes.
\end{description}

Thus, the continuum fitting process is iterative and a satisfactory continuum was obtained after $W_{total}$=11 iterations. The chosen window sizes for each iteration $x$ were

\begin{eqnarray}
&W_{width} = &\left\{
\begin{array}{c l}
    2^{x} \ \ \ \ \ \ \ \ \ ; & 1\le x \le6\\
    2^{11-(x-1)}; & 7\le x \le11
\end{array}\right.
\end{eqnarray}
and the $W_{pass}$ for each $x$ were

\begin{eqnarray}
&W_{pass} = &\left\{
\begin{array}{c l}
    N_{pix}; & 1\le x \le6\\
    \frac{N_{pix}}{11- (x-1)}; & 7\le x \le11
\end{array}\right.
\end{eqnarray}

The pseudo-continuum fitting works similarly to the TM filter but uses the median value of the window around the desired pixel instead of averaging for slightly modified window parameters.

\begin{figure*}
\centering
\includegraphics[width=0.9\textwidth, height=0.83\textwidth,clip=]{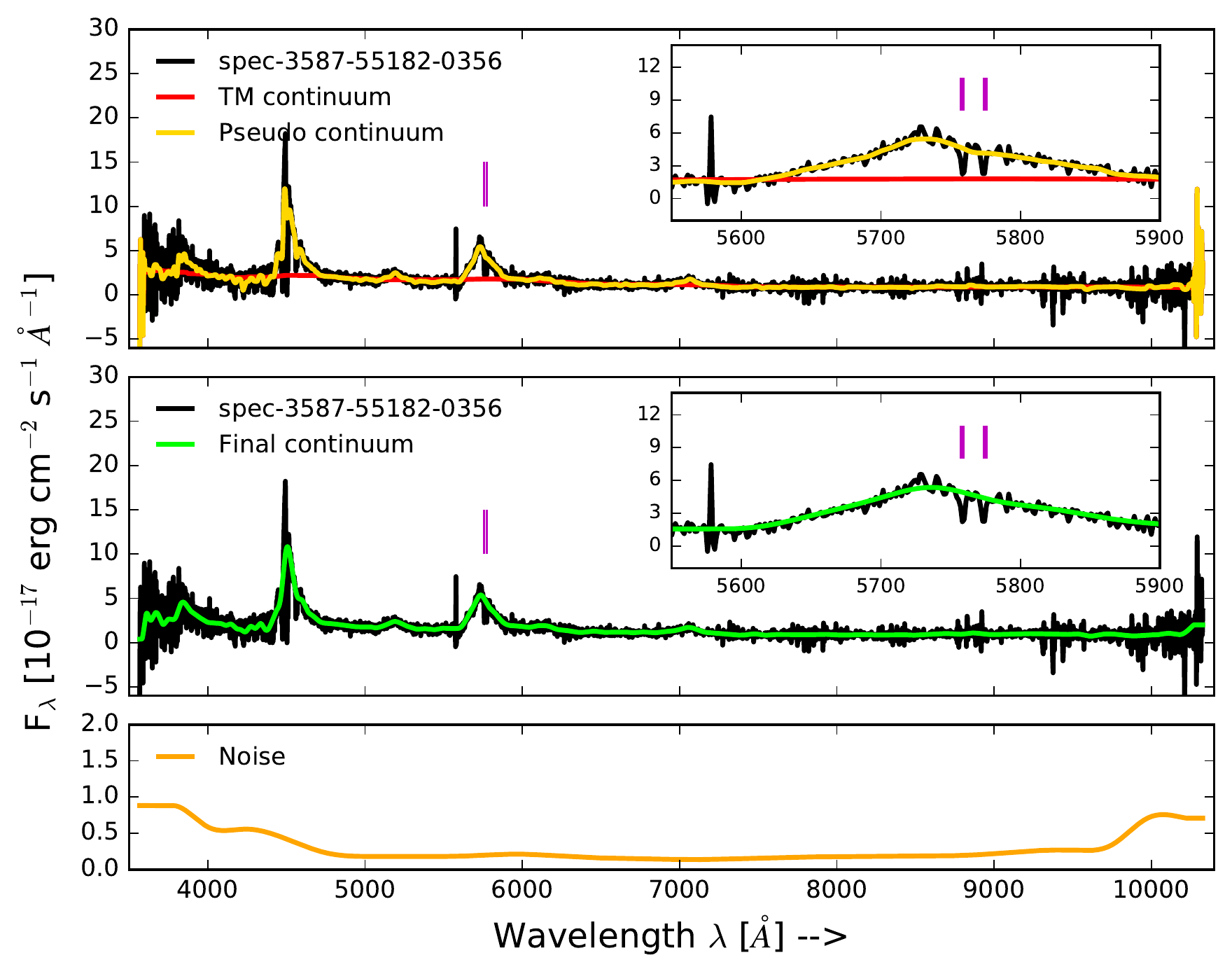}
\caption{The top panel shows the quasar spectrum (black), the derived TM (red), and pseudo (yellow) continua. The final continuum (green) is over-plotted on the data in the middle panel. Also shown in the plots as the vertical magenta lines near \mbox{$\lambda$=5780 \AA} is a \mbox{Mg{\sc\ II}} doublet at \mbox{z$_{2796}$=1.059} identified in the current work. The bottom panel shows the estimated noise spectrum using the median method for the same quasar spectrum.}
\label{fig_continuum_example}
\end{figure*}

The final combined continuum using TM and median filter technique was obtained as follows. Since the prime focus is to trace the emission lines, a window of width $\sim$30 pixels, just enough to include the narrowest emission line, was adopted. The window was moved over the entire spectrum and at every step, the average values of the continuum provided by the TM and median algorithms in the window were compared. In the regions where an emission line is present, the median values will follow the varying flux more closely than the mean values. Therefore in the windows (spectral regions) where the median values are larger than the mean values, an emission line is being traced. As the tracing of the emission lines is used in conjunction with the bona-fide TM filter continuum, we call the former the pseudo-continuum. The difference between the final and the TM filter continua is shown as a waterfall plot in Fig. \ref{fig_con_diff} for 10000 randomly selected non-BAL DR12Q spectra. Like expected, the maximum difference between the continuum values occurs near and within the presence of the broad emission lines (Ly-$\alpha$, \mbox{Si{\sc\ IV}+O{\sc\ IV}}, \mbox{C{\sc\ IV}}, \mbox{C{\sc\ III]}}, and \mbox{Mg{\sc\ II}}) as marked in the figure.

Fig. \ref{fig_continuum_example} shows an example SDSS DR12Q spectrum. The top panel shows the quasar spectrum (black), the derived TM (red), and pseudo (yellow) continua. The final continuum (green) is over-plotted on the data in the middle panel. Also shown in the plots as vertical magenta lines near \mbox{$\lambda$=5780 \AA} is a \mbox{Mg{\sc\ II}} doublet at \mbox{z$_{2796}$=1.059} identified in the current work. The inset plot is a zoomed version of the region near the \mbox{Mg{\sc\ II}} doublet to show the quality of the continuum fitting.

The noise estimation was performed by splitting the spectrum into several distinct blocks of 500 pixels. The first and the last blocks were sub-divided into narrower blocks of 100 pixels each, to account for the high noise levels at the start and end of the SDSS spectrum. A cubic spline interpolation then interpolates the median flux value of each of the blocks to get the final noise spectrum. The bottom panel in Fig. \ref{fig_continuum_example} shows an example of the noise estimation. The first and the last 30 pixels of the spectrum were ignored for both the continuum and noise estimation.

\subsection{The doublet finder}
The doublet finder scans the spectra to get the candidate list of \mbox{Mg{\sc\ II}} doublets. The detections with the integrated signal-to-noise SNR (Eq. \ref{eq_int_snr}) of three or more for each line of the doublet and with a wavelength separation matching the \mbox{Mg{\sc\ II}} doublets ($7.1 \pm 0.25$ \AA\ in the rest-frame) in the redshift range $0.35 \le z_{2796} \le 2.3$, were retained.
\begin{eqnarray}
\label{eq_int_snr}
SNR_{2796}=\sum_{i=p_{1}}^{p_{2}} \left( \frac {C_i-F_i} {C_i} \right)/ \left(\sum_{i=p_{1}}^{p_{2}}\sigma_i^{2}\right)^{\frac{1}{2}}
\end{eqnarray}
where $F$ is the flux, $C$ is the continuum, $\sigma$ is the noise, and $p_1$ and $p_2$ represent the starting and ending pixels\footnote{The first and the last pixel where the spectrum has a flux deficit compared with the continuum.} of the line. To this initial list, we applied a cut of $SNR \ge 6.0, 3.0$ respectively for $\lambda 2796, 2803$ lines of the doublet. The preliminary catalogue contained 74,550 absorption system candidates.

The lower and upper limits of the search window for a given spectrum were defined by the location of the $Ly\alpha$\footnote{To avoid false \mbox{Mg{\sc\ II}} detections in the $Ly\alpha$ forest region.} and \mbox{Mg{\sc\ II}}\footnote{Redshift of the intervening \mbox{Mg{\sc\ II}} absorbers must be lower than the redshift of the quasar.} emission. These limits were modified if the location of the \mbox{Mg{\sc\ II}} doublet fell outside the wavelength range of the SDSS spectra. The precise definition of the search window is given below

\begin{eqnarray}
\label{eq_search_window}
&z_{min} = &max(0.35, z_{QSO}+\Delta z_{Ly-\alpha}) \\
&z_{max} = &min(2.3, z_{QSO}-\Delta z_{QSO})
\end{eqnarray} where $\Delta z_{Ly-\alpha}=0.1$ corresponds to a velocity separation of \mbox{v $\ge$ 30000 km s$^{-1}$} from the $Ly\alpha$ emission and $\Delta z_{QSO}=0.03$ corresponds to \mbox{v $\le$ 9000 km s$^{-1}$} from the \mbox{Mg{\sc\ II}} emission. The quasar redshift $z_{QSO}$ corresponds to the $z_{PIPE}$ field of the DR12Q catalogue.

The redshifts of the two lines of the doublet were measured by fitting two Gaussians, one for each member of the doublet. Although, the spectral shape of the \mbox{Mg{\sc\ II}} doublets is best represented by Voigt profiles (convolution of the Gaussian and the Lorentzian profiles), the use of Gaussians suffice for the current work as they are used for the redshift estimation only. The rest-frame EW ($W_r$) of the line is measured using the original spectrum as:

\begin{equation}
W_r = \frac{1}{1+z_{2796}}\ \sum_{i=p_{1}}^{p_{2}} \left( \frac {C_i-F_i} {C_i} \right) \Delta \lambda
\label{eq_ew}
\end{equation} where $\Delta \lambda$ is the pixel resolution in \r{a}ngstr\"{o}ms. The corresponding error $\sigma_W{_r}$ is 
\begin{equation}
\sigma_{W_r} =\frac{1}{1+z_{2796}}\ \left (\sum_{i=p_1}^{p_2} \left[\frac {\sigma_{i}} {C_{i}} \Delta \lambda \right]^{2}\right)^{1/2}
\label{sigma_ew}
\end{equation} where $\sigma$ is the pixel noise estimated as explained in the previous section.

\section{Analysis}
\subsection{Catalogue refinement}
\label{sec_cat_ref}
\subsubsection{Cut 1 - OH band cuts, \mbox{Ca{\sc\ II}}, and \mbox{C{\sc\ IV}} systems}
The first set of cuts were to eliminate contamination from the sky lines and the lines arising from the Galactic absorption. The SDSS spectra can contain strong sky lines such as [OIII], [OI], Na, and Hg lines. These are handled separately and explained in the next section \ref{sec_strong_sky_emission_line}. Other than the above listed strong sky lines, the SDSS spectra also contain the calcium \mbox{Ca{\sc\ II} ($3934, 3969$ \AA)} arising from our Galaxy. We ignored 100 possible \mbox{Ca{\sc\ II}} systems by applying a 60 \AA\ mask ($3920 \le \lambda_{2796} \le 3980\ \AA$). There are also numerous other sky line artefacts mimicking \mbox{Mg{\sc\ II}} doublet in the OH band \mbox{($\lambda \ga 6900$ \AA)} arising primarily due to the difficulty of sky subtraction near these wavelengths in the SDSS spectra. These are the sharp black peaks in Fig. \ref{fig_wave_dist_sky}. We used the SDSS bitmasks \citep{bolton2012} to identify the bad pixels in the OH band and detections within 4 \AA\ from a masked SDSS pixel were removed from the catalogue. The gaps in the green histogram in Fig. \ref{fig_wave_dist_sky} show the result of this cut (27 per cent) in the OH band. 

Finally, there is also a chance of erroneously identifying a \mbox{C{\sc\ IV}} as a \mbox{Mg{\sc\ II}} doublet because of the difficulty in distinguishing the two. The \mbox{C{\sc\ IV}} doublets arise from a differently distributed population of objects as compared to the \mbox{Mg{\sc\ II}}. However, the \mbox{C{\sc\ IV}} systems can also help in tracing the galaxies and the LSSs. Subsequently we flagged all the possible \mbox{C{\sc\ IV}} systems (the systems detected below $\Delta z = 0.03$ from \mbox{C{\sc\ IV}} emission) but did not remove them from the catalogue. Note that the flagged systems ($\sim$21 per cent) were not included in calculations of the evolution of the number density of the \mbox{Mg{\sc\ II}} systems, and the null tests (section \ref{stats_evolution_mgii}, and \ref{sec_null_tests}).

\subsubsection{Cut 2 - Strong sky emission lines}
\label{sec_strong_sky_emission_line}

\begin{figure}
\centering
\includegraphics[width=0.45\textwidth, height=0.33\textwidth,clip=]{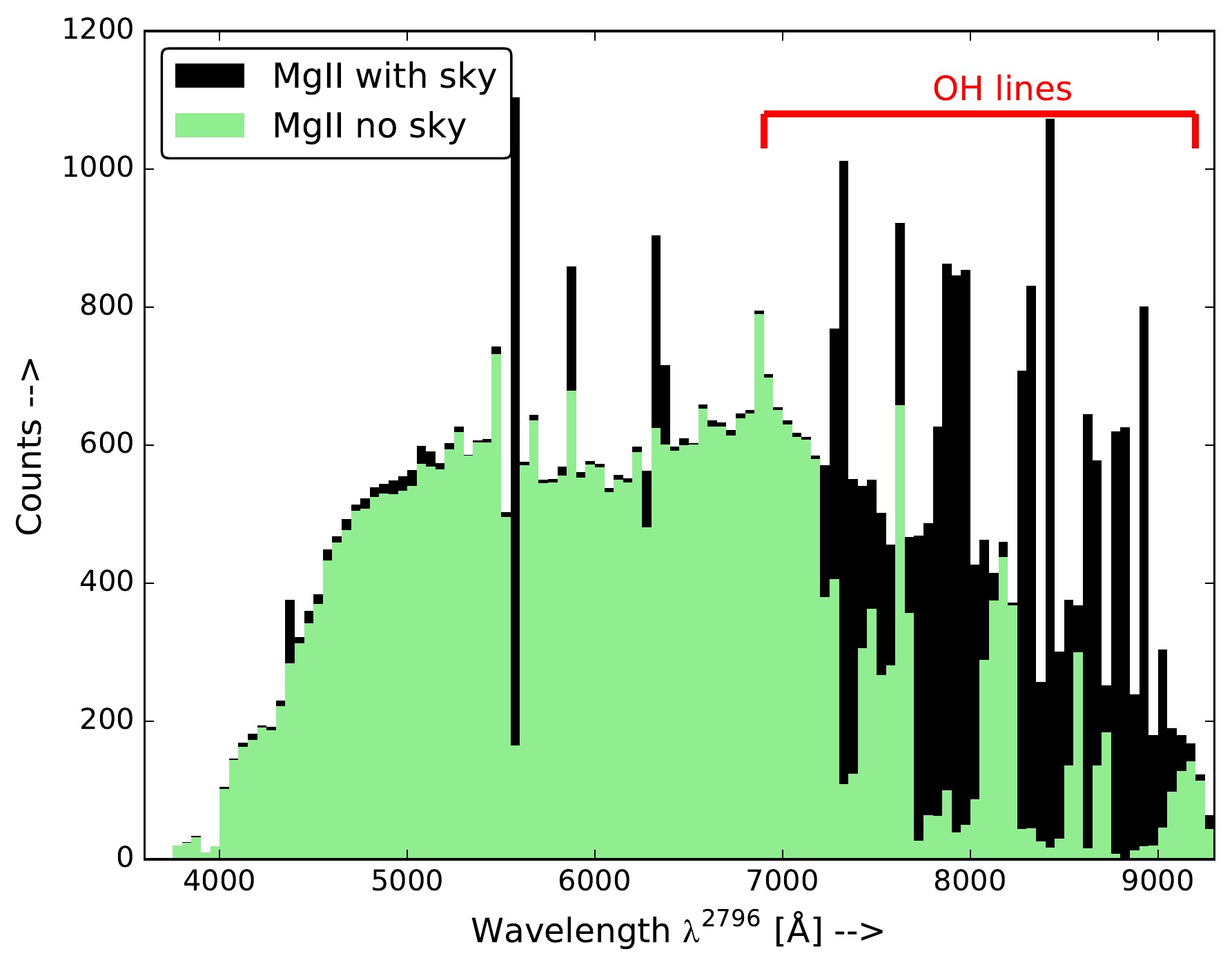}
\caption{Wavelength distribution of the \mbox{Mg{\sc\ II}} catalogue before (black) and after (green) applying the SDSS bitmasks, and results of the SF algorithm. The sharp black peaks near the sky line wavelengths as mentioned in the text represent the original catalogue contamination due to the sky lines.}
\label{fig_wave_dist_sky}
\end{figure}

The SDSS quasar spectra contain strong night sky emission lines at \mbox{4364 \AA\ ([OIII]), 5578 \AA\ ([OI])}, \mbox{5895 \AA\ (Na)}, \mbox{6302 \AA\ ([OI])}, \mbox{6365 \AA\ ([OI])}, and \mbox{7246 \AA\ (Hg)} (e.g. \citet{massey2000}) leading to an underestimation of the noise near the sky lines causing spurious \mbox{Mg{\sc\ II}} detections. Since many of these lines are not captured by the SDSS BITMASK we perform an additional strong sky line refinement. The sharp black peaks near the above mentioned wavelengths in Fig. \ref{fig_wave_dist_sky} shows the contamination in the catalogue. A simple solution would be to apply narrow masks at these wavelengths. But, some of these lines are attributed to auroral activities and not all quasar spectra contain them. Hence we developed a separate sky line finder (SF) to reduce the potential contamination of our absorber catalogue due to the sky lines. The SF scans each quasar spectrum to construct individual narrow sky line masks. 

\begin{figure}
\includegraphics[width=0.48\textwidth, height=0.5\textwidth,clip=]{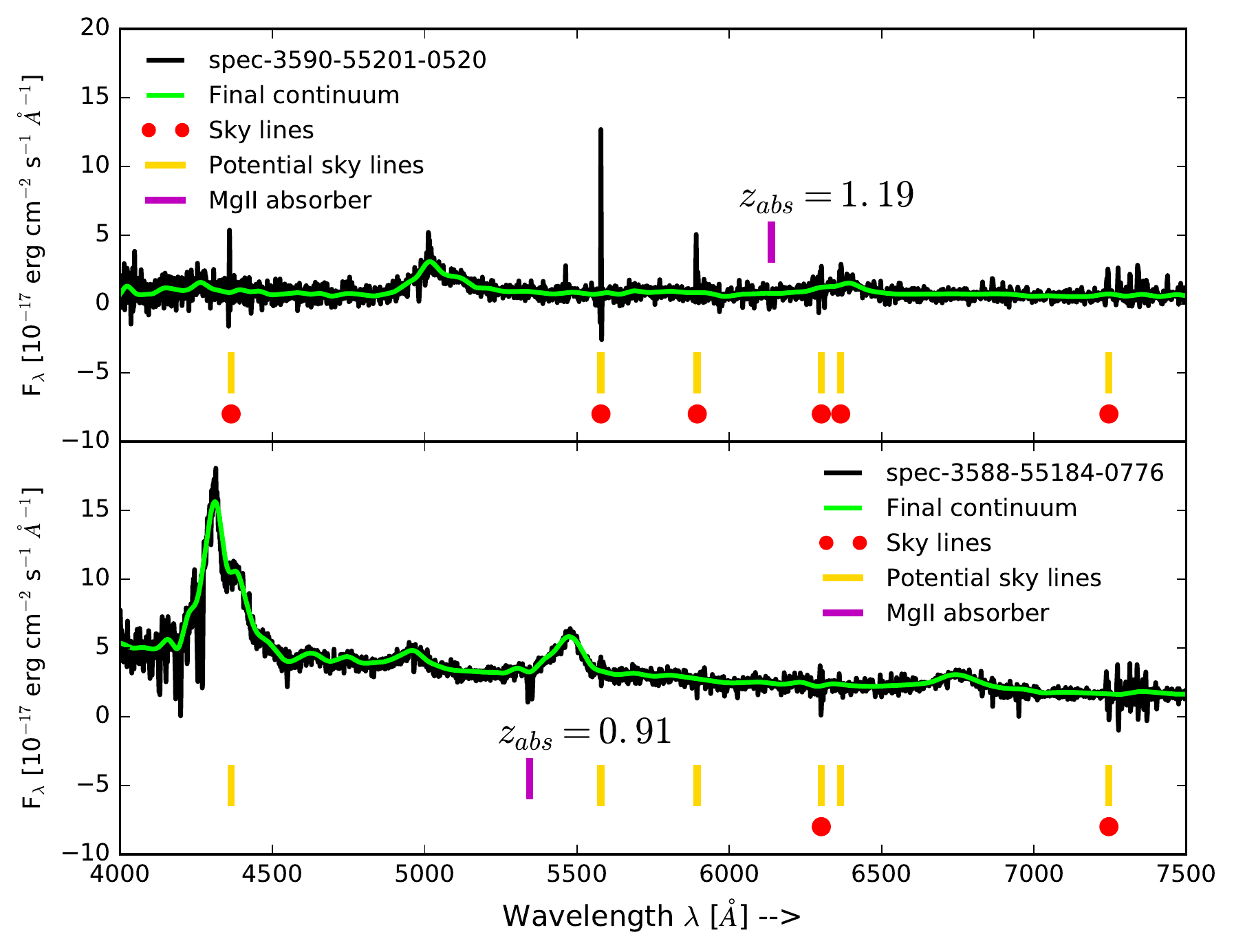}
\caption{Strong sky line finder algorithm acting on the spec-3590-55201-0520 (spec-3588-55184-0776) quasar spectrum in the top (bottom) panel. The yellow lines show the region where the sky lines were searched for, red circles show the strong sky lines present and captured by the algorithm. The top panel is an example where the sky lines affect the data quality significantly and are all picked by the SF finder. The bottom panel shows a cleaner spectrum where the SF finder only picks two lines. The \mbox{Mg{\sc\ II}} doublet at $z_{2796}$=1.2 (0.9) in the top (bottom) panel picked by the doublet finder is also marked using the magenta line.}
\label{fig_strong_sky_line}
\end{figure}

The SF algorithm works as follows. We chose three windows for the [OIII], [OI], Na, and Hg lines such that the central window completely covers the sky line with control windows on either side. The size of the central window varied between 5 and 20 pixels depending on the pixel width of the line while the control windows always covered five pixels each. The scatter in the flux level in all the three windows was calculated and the spectrum was flagged for a particular line if the scatter in the central window was $\sigma_{central} \ge 3.5\ \sigma_{control}$. 

An example of the SF process is shown in Fig. \ref{fig_strong_sky_line} for two quasar spectra. The yellow vertical lines show the location of the potential sky lines and the presence of the red circle indicates that the quasar spectrum is affected by the corresponding sky line. The bottom panel is a cleaner spectrum with only two strong sky lines picked by the SF finder (red circles) compared to the top panel which contains five strong lines. The \mbox{Mg{\sc\ II}} absorption system at $z_{2796}$=1.2, and 0.9 picked by the doublet finder is marked as the vertical magenta line in the two panels. Our absorber catalogue was cross-correlated with the strong sky line catalogue from the SF algorithm. The absorbers present near the sky line (central window) wavelengths in the spectra of flagged quasars were subsequently removed. The sky line catalogue obtained using the SF algorithm is also available online along with the \mbox{Mg{\sc\ II}} catalogue.

The SF cut removed 2.5 per cent of the systems near the strong sky lines. The green histogram in the Fig. \ref{fig_wave_dist_sky} show the distribution of the absorption wavelengths after incorporating the results of the SF finder and the SDSS BITMASK.

\subsection{False detections and caveats}
In this section we caution the reader about possible false detections of \mbox{Mg{\sc\ II}} absorption systems and other systematic effects. We estimate less than eight per cent of the detections in our catalogue as false detections due to artefacts of the continuum fitting. The underestimation of the emission lines by the pseudo-continuum could lead to false detections near the strong emission lines like \mbox{C{\sc\ III}]}, and \mbox{C{\sc\ IV}} (see Fig. \ref{fig_con_diff}). The artefacts could also be due to the overestimation of the continuum near the pixels where the pseudo-continuum takes over the TM filter continuum. The accurate method of estimating the detection errors due to the continuum artefacts is using mock spectra. But, we ignore this step as the catalogue contains less than eight per cent of the systems near strong emission lines. These systems were removed from the catalogue. 

As mentioned already, we flagged the possible ($\sim$21 per cent) \mbox{C{\sc\ IV}} ($\Delta z = 0.03$ from \mbox{C{\sc\ IV}} emission) absorbers but did not remove them from the catalogue since they could be useful for other LSS studies.

To estimate the percentage of false detections in the catalogue, we used random Gaussian-noise only simulations. We chose to use simulations instead of visual checks to avoid any observer bias. For the noise only simulations we randomly picked 45,000 from the original 266,433 searched quasars. The continua derived using the original spectra were used as templates and separated into four blocks ($\lambda:\ \le 4000,\ 4000\ -\ 6200,\ 6200\ -\ 7500,\ \ge 7500\ \AA $) for each spectra. Random Gaussian-noise with zero mean and sigma equal to the standard deviation of the spectra in the respective block was added to this template. The noise only simulations are then processed through the doublet finder and the  obtained catalogue was refined using the same procedure described in the previous section. The refined catalogue contains 516 systems detected in 45,000 noise only simulations. Projecting this false detection rate,  we estimate $\sim$3055 false detections in the overall catalogue. Thus, less than eight percent of the systems in our catalogue could be due to noise. 

\section{Results}
\subsection{The \mbox{Mg{\sc\ II}} catalogue}
\label{catalogue_description}

\begin {table*}
\centering
\caption {An excerpt from the \mbox{Mg{\sc\ II}} absorption catalogue. The columns represent: the SDSS name of the quasar spectra; plate-mjd-fibre numbers of the spectra; RA, Dec. (J2000); quasar redshift; quasar PSF $i$-band magnitude; absorption redshift and the error; rest-frame EW of the two lines of the doublet along with their errors; the flag showing if the detection could be a \mbox{C{\sc\ IV}} line; $SNR_{CON}$ calculated in the current work (see section \ref{sec_null_tests}). The full catalogue is available online.}

\resizebox{\textwidth}{!}{\begin {tabular} {cccccccccccccc}
\hline
SDSS name & plate-mjd-fibre & RA & Dec. & z$_{QSO}$ & PSF $i$ & $z$ & $\delta z$ & W$_1$ & $\delta$W$_1$ & W$_2$ & $\delta$ W$_2$ & \mbox{C{\sc\ IV}} & SNR$_{con}$\\
& & [Deg.] & [Deg.] & & mag & & [$10^{-4}$] & [\AA] & [\AA] & [\AA] & [\AA] & flag\\
\hline\\
003304.79+004338.1 & 3587-55182-0602 & \ 8.269998 & \ 0.727270 & 2.217 & 19.585 & 1.362 & 4.871 & 1.334 & 0.180 & 0.760 & 0.146 & 0 & 12.353\\
003416.61+002241.1 & 3587-55182-0691 & \ 8.569213 & \ 0.378083 & 1.627 & 17.290 & 1.188 & 0.340 & 1.134 & 0.060 & 1.027 & 0.063 & 0 & 29.756\\
003640.34+011449.4 & 3587-55182-0714 & \ 9.168105 & \ 1.247081 & 2.348 & 20.204 & 1.369 & 2.049 & 2.591 & 0.193 & 1.583 & 0.164 & 0 & 10.333\\
004034.19+001356.4 & 3588-55184-0776 & 10.142481 & \ 0.232342 & 2.543 & 20.348 & 0.909 & 1.430 & 1.556 & 0.123 & 1.307 & 0.119 & 1 & 11.015\\
003746.95-002052.5 & 3587-55182-0222 & \ 9.445635 & -0.347941 & 2.617 & 20.836 & 0.802 & 3.969 & 0.838 & 0.129 & 0.484 & 0.117 & 1 & \ 6.717\\
003746.95-002052.5 & 3587-55182-0222 & \ 9.445635 & -0.347941 & 2.617 & 20.836 & 1.524 & 4.974 & 1.701 & 0.254 & 1.017 & 0.202 & 0 & \ 6.717\\
003542.19-011313.2 & 3587-55182-0282 & \ 8.925794 & -1.220347 & 2.185 & 20.061 & 0.896 & 1.309 & 0.863 & 0.100 & 0.839 & 0.109 & 0 & 10.910\\
003432.72-004108.4 & 3587-55182-0344 & \ 8.636340 & -0.685670 & 1.414 & 20.989 & 0.777 & 6.594 & 1.896 & 0.230 & 0.784 & 0.202 & 0 & \ 4.438\\
004439.24-004207.2 & 3589-55186-0218 & 11.163518 & -0.702013 & 2.060 & 21.773 & 0.754 & 2.721 & 3.272 & 0.536 & 1.365 & 0.390 & 0 & \ 1.682\\
005058.61-005845.6 & 3590-55201-0122 & 12.744232 & -0.979337 & 2.537 & 20.399 & 1.922 & 4.384 & 2.873 & 0.341 & 2.752 & 0.429 & 0 & \ 6.558\\
... & ... & ... & ... & ... & ... & ... & ... & ... & ... & ... & ... & ... & ...\\
... & ... & ... & ... & ... & ... & ... & ... & ... & ... & ... & ... & ... & ...\\
\\\hline
\end {tabular}}
\label{mgii_catalogue_table}
\end {table*}

\begin{figure}
\includegraphics[width=0.48\textwidth, height=0.4\textwidth,clip=]{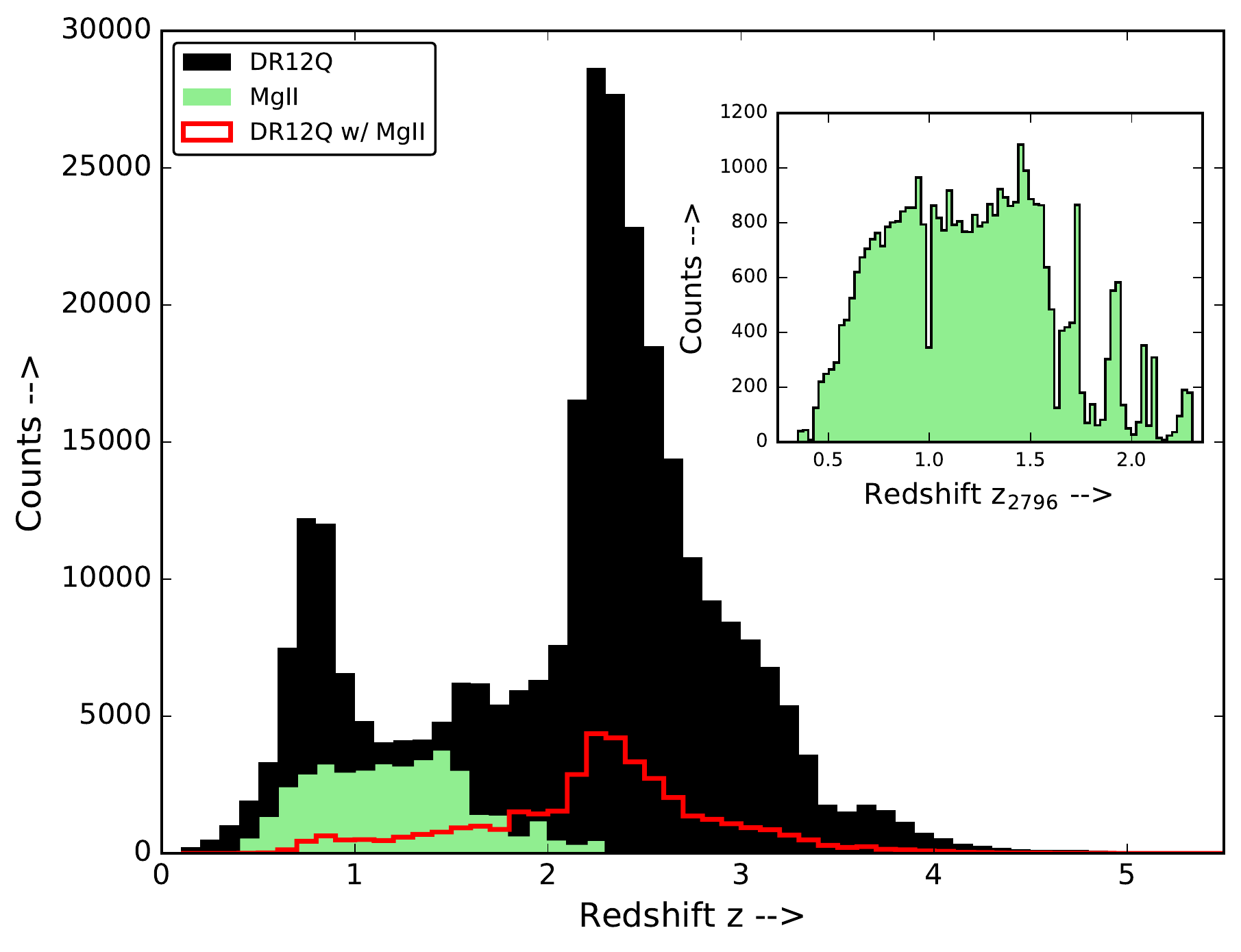}
\caption{The redshift distribution of the catalogued \mbox{Mg{\sc\ II}} absorption (green) systems and the background quasars (black). The red stepped histogram represents the redshift distribution of the quasars hosting the \mbox{Mg{\sc\ II}} absorbers in their spectra. The histogram is binned with \mbox{$\Delta z=0.1$}. The horizontal axis is limited to \mbox{z = 5.5} for clarity, but the most distant background quasar is at \mbox{z = 7.1}. The inset plot shows the redshift distribution of the absorbers with a finer bin, \mbox{$\Delta z=0.025$}.}
\label{fig_z_abs_qso}
\end{figure}

\begin{figure*}
\centering
\includegraphics[width=0.9\textwidth, height=0.8\textwidth,clip=]{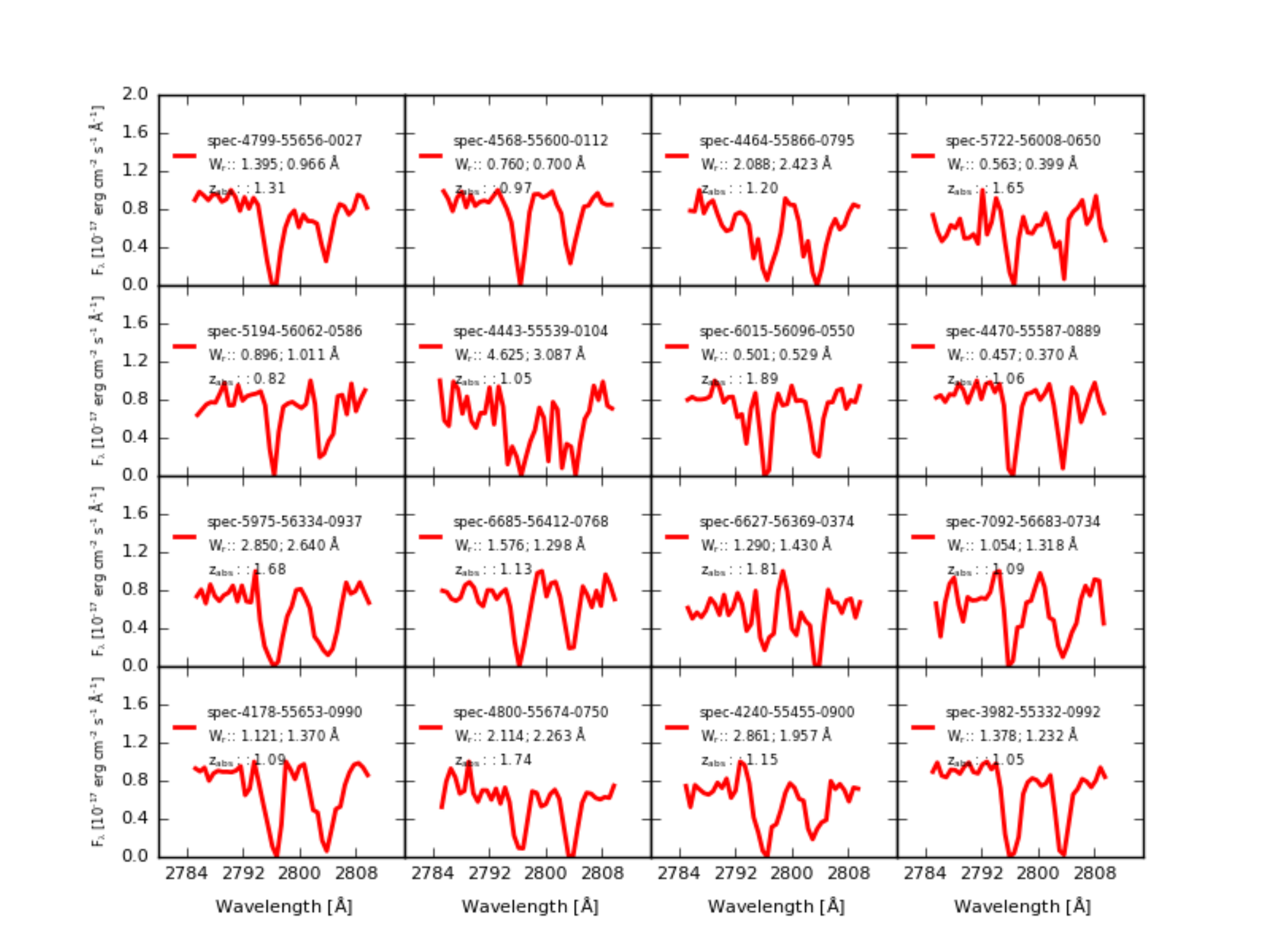}
\caption{A random selection of 16 \mbox{Mg{\sc\ II}} absorption systems detected in the current work. The absorbers are shown in their rest-frame wavelengths with the flux normalised to one. The quasar spectrum name ``spec-PLATE-MJD-FIBER'' along with the EW of the two lines, and the redshift are given in each panel.}
\label{fig_mgii_random}
\end{figure*}

Table \ref{mgii_catalogue_table} shows an excerpt from the \mbox{Mg{\sc\ II}} catalogue produced in this work\footnote{The catalogues produced in this work can be downloaded from \url{http://srini.ph.unimelb.edu.au/mgii.php}.}. The resultant catalogue after all the above mentioned cuts contains 39,694 absorbers in the EW\footnote{From now on, the W$_r$ will always mean the rest-frame EW calculated for the $\lambda$2796 line unless otherwise specified.} range \mbox{$0.2 \le W_{r} \le 6.2$ \AA}, constrained to a redshift range \mbox{$0.35 \le z_{2796} \le 2.3$}. Fig. \ref{fig_z_abs_qso} shows the observed redshift distributions of the \mbox{Mg{\sc\ II}} absorbers (green) and the background quasars (black). The red histogram corresponds to the redshift distribution of the quasars that host \mbox{Mg{\sc\ II}} in their spectra. The inset plot shows the redshift distribution of the absorbers with a finer bin, \mbox{$\Delta z=0.025$}. The dips near z=0.41 and z=1.0 correspond to the removal of \mbox{Ca{\sc\ II}} systems and the \mbox{$5577$ \AA} [OI] line. A random selection of 16 \mbox{Mg{\sc\ II}} doublets detected in this work is shown in Fig. \ref{fig_mgii_random}.

\begin{figure}
\centering
\includegraphics[width=0.5\textwidth, height=0.4\textwidth,clip=]{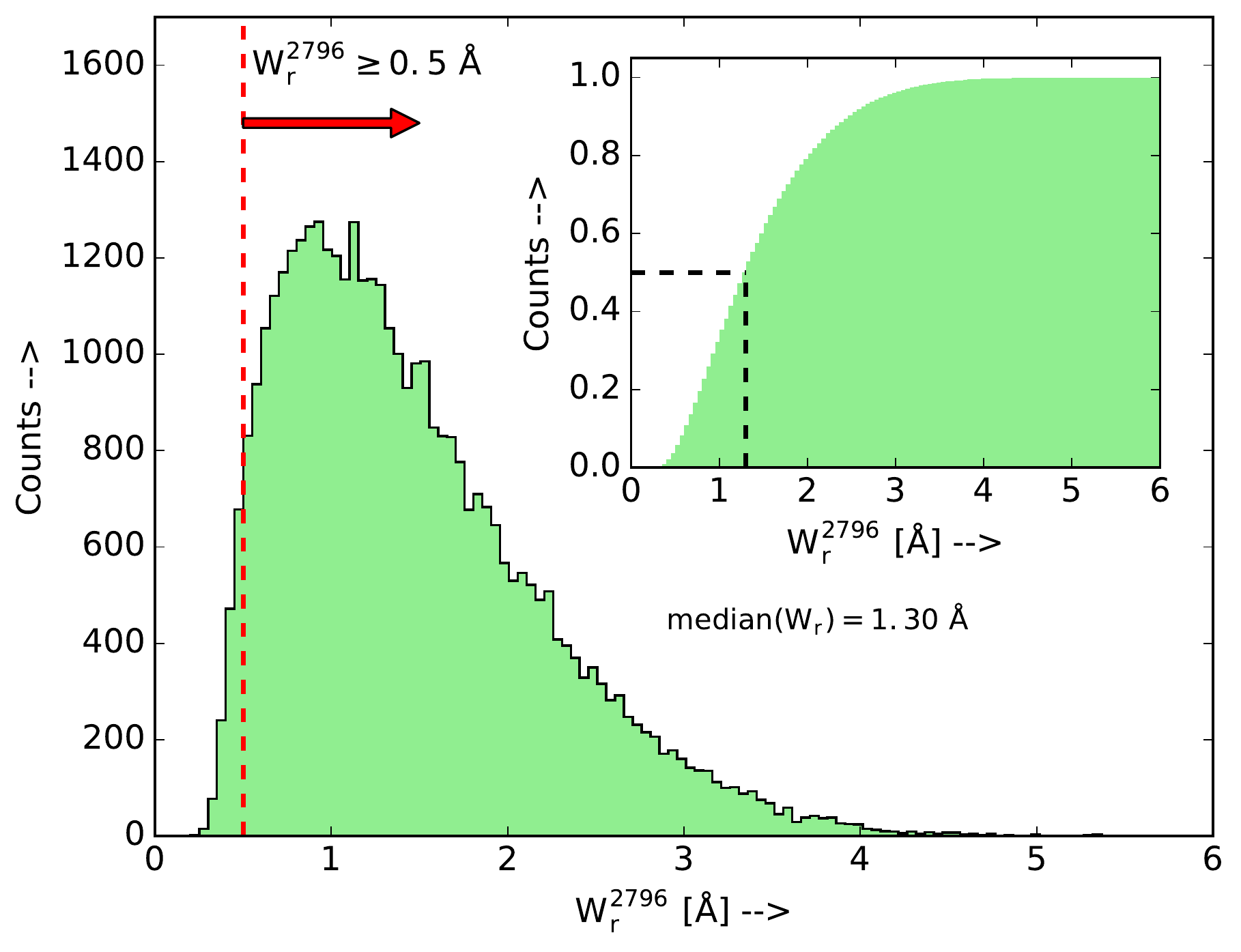}
\caption{The rest-frame EW distribution of the catalogued \mbox{Mg{\sc\ II}} absorption binned with $\Delta$W$_r$=0.05 \AA. The red dashed line marks the transition point from mild to strong systems at \mbox{W$_r$ = 0.5 \AA}. The distribution peaks around 0.9 \AA\ with a median value of $\widetilde W_r$ = 1.3 \AA,\ which is marked in the cumulative distribution in the inset plot.}
\label{fig_ew_dist}
\end{figure}

\begin{figure}
\centering
\includegraphics[width=0.46\textwidth, height=0.4\textwidth,clip=]{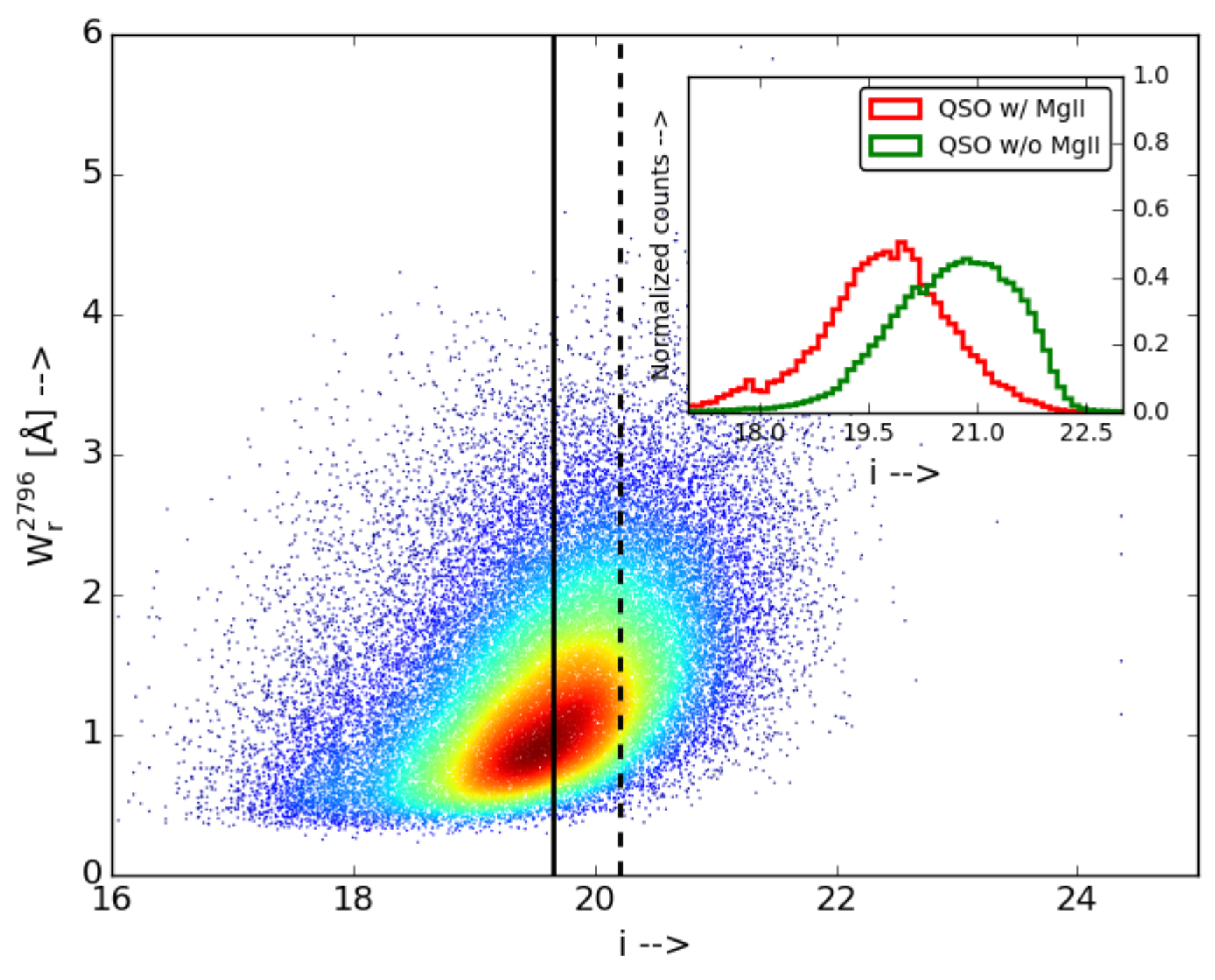}
\caption{The dependence of the \mbox{Mg{\sc\ II}} detection on the brightness of the background quasar. Weak \mbox{Mg{\sc\ II}} systems are predominantly detected in brighter quasars as expected. The black solid line corresponds to the median value of $i$ magnitude (19.65) for the quasars hosting the \mbox{Mg{\sc\ II}} doublets detected in this work. The inset plot shows the distribution of the $i$ magnitude for the quasars with (without) the \mbox{Mg{\sc\ II}} absorption systems in red (green). The drop in the red histogram near $i=20.2$ corresponds to the DR12Q bright target selection criteria \citep{richards2002, alam2015}. This limit is marked with the black dashed line in the density plot.}
\label{fig_app_mag_ew}
\end{figure}

The EW distribution of the catalogued \mbox{Mg{\sc\ II}} absorption is shown in Fig. \ref{fig_ew_dist}. Close to 28 per cent of the systems lie in the range \mbox{$0.5\ <\ $W$_{r}\ \le$ 1.0 \AA} and 21 per cent of the systems are extremely strong with \mbox{W$_{r}>$ 2.0 \AA}. The systems with \mbox{1.0 $\le$ W$_{r} < 2.0$ \AA} account for 47 per cent of the catalogue. Fewer than one per cent of absorbers in the catalogue are \emph{Weak absorbers} with \mbox{W$_{r} \le 0.3$ \AA}. Normally, the transition from weak to strong \mbox{Mg{\sc\ II}} absorbers happens near \mbox{0.3 \AA}. However, considering the difficulties in accurately measuring such small EWs in low resolution spectra like the SDSS, we classify the absorbers with \mbox{W$_{r}\ < 0.5\ \AA$} as \emph{Mild absorbers}. The mild absorbers account for four per cent of the catalogued systems (left of the red dashed line in the Fig. \ref{fig_ew_dist}). The reason for such a low detection of weak systems is due to the dependence of the \mbox{Mg{\sc\ II}} detection on the brightness of the background quasar. The density plot in Fig. \ref{fig_app_mag_ew} shows the relationship between the \mbox{$i$-band} magnitude of the quasar and the EW of the detected absorbers. As expected, and noted in other similar studies \citep{nestor2005,quider2011}, it is evident from the figure that weaker \mbox{Mg{\sc\ II}} systems are detected predominantly in brighter quasars. The black dashed line at $i=20.2$ in the density plot corresponds to the target selection $i$ magnitude limit for the SDSS main survey \citep{richards2002, alam2015}. The $i=20.2$ limit can also be seen as a drop in the red histogram in the inset plot of Fig. \ref{fig_app_mag_ew}. The red (green) histogram in the inset plot shows the $i$ magnitude distribution of the quasars with (without) the \mbox{Mg{\sc\ II}} systems detected in their spectra. More than 66 per cent of the DR12Q are fainter than this limit resulting in a low detection of the weak \mbox{Mg{\sc\ II}} systems.

\begin{figure}
\centering
\includegraphics[width=0.47\textwidth, height=0.4\textwidth,clip=]{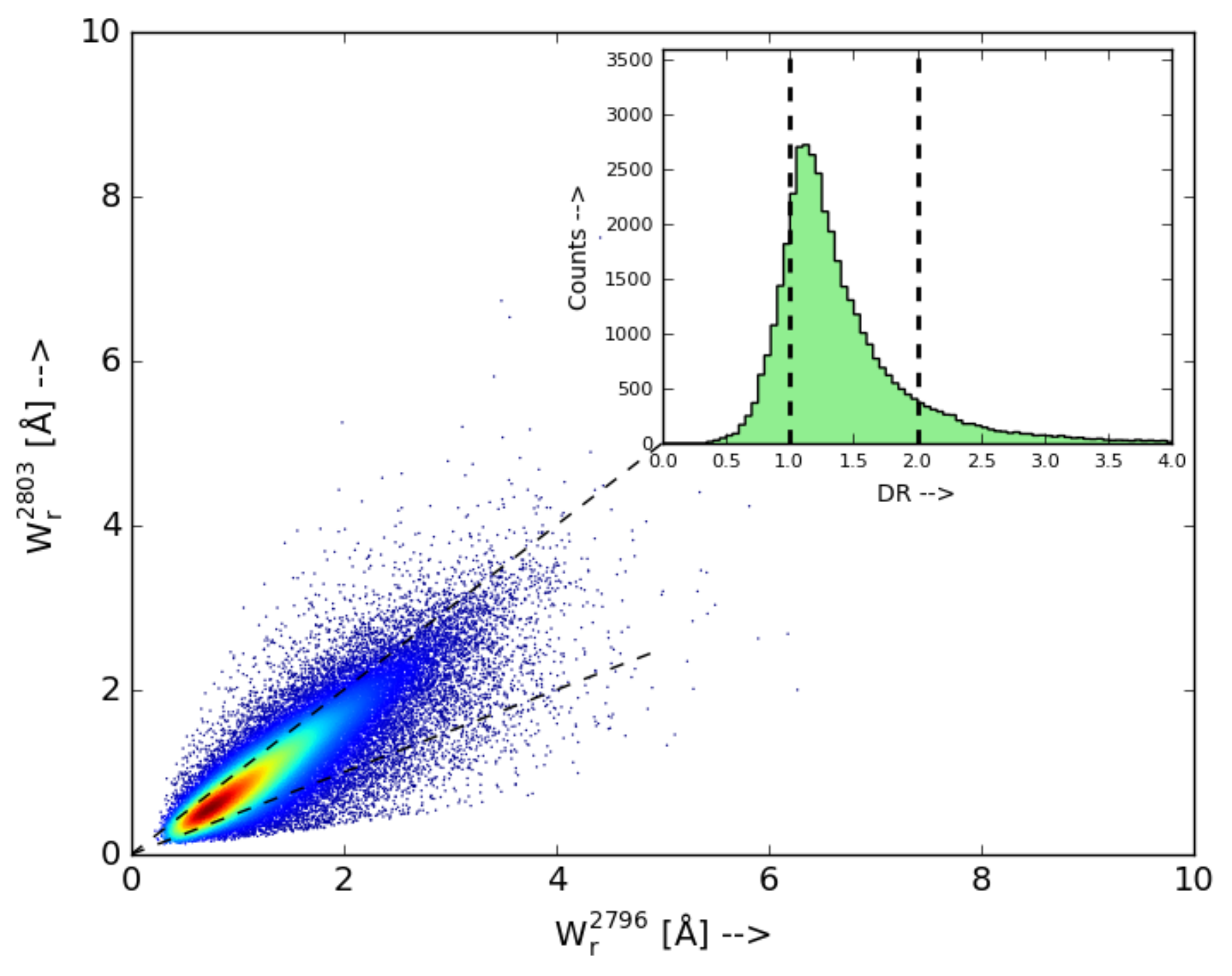}
\caption{The relationship between the line strengths of the two lines of the \mbox{Mg{\sc\ II}} doublet. Majority of the catalogued systems lie within the theoretical limits shown as black dashed lines. See the text for more details.}
\label{fig_ew_dr}
\end{figure}

The relationship between the line strengths of the two lines of the \mbox{Mg{\sc\ II}} doublet is shown in Fig. \ref{fig_ew_dr}. The ratio of the line strengths (doublet ratio), should ideally lie in the range one (completely saturated), and two (unsaturated). Due to associated errors in the EWs, roughly 29 per cent of the systems lie outside the ideal doublet ratio range \mbox{(17 per cent $<$ 1.0 and 12 per cent $>$ 2.0)}. The doublet ratio distribution is shown as the histogram in the inset plot in Fig. \ref{fig_ew_dr}. Our doublet ratio measurements are inclined towards a value of 1.0 rather than 2.0 indicating that majority of the detected systems are saturated. For saturated systems, the blending of the two lines of the doublet could also introduce errors in the EW measurements explaining why a majority of systems outside the theoretical doublet ratio limit appear below the lower end (17 per cent with doublet ratio $<$ 1.0).

\subsection{Catalogue comparison}
\label{cat_comp}
To determine the quality of the catalogue we performed a comparison of our catalogue to the publicly released ZM13\footnote{The \mbox{Mg{\sc\ II}} catalogue detected from the DR12Q was downloaded from \url{http://www.guangtunbenzhu.com/\#!jhu-sdss-metal-absorber-catalog/acrow}} catalogue. The ZM13 \mbox{Mg{\sc\ II}} DR12 catalogue contains 41,895 absorption systems detected at a SNR level of 4.0 and 2.0 for the two lines respectively. 

The ZM13 searched for absorption lines in the spectra of 57,479 DR12 quasars. Of this, we only used absorption systems detected in the spectra of 47,550 DR12Q for the catalogue comparison purpose. The ignored 9929 quasars are either marked as BAL quasars (5915 quasars) in the DR12Q catalogue or not included (4014 quasars) in the official DR12Q release \citep{paris2015}.

\begin{figure}
\centering
\includegraphics[width=0.48\textwidth, height=0.4\textwidth,clip=]{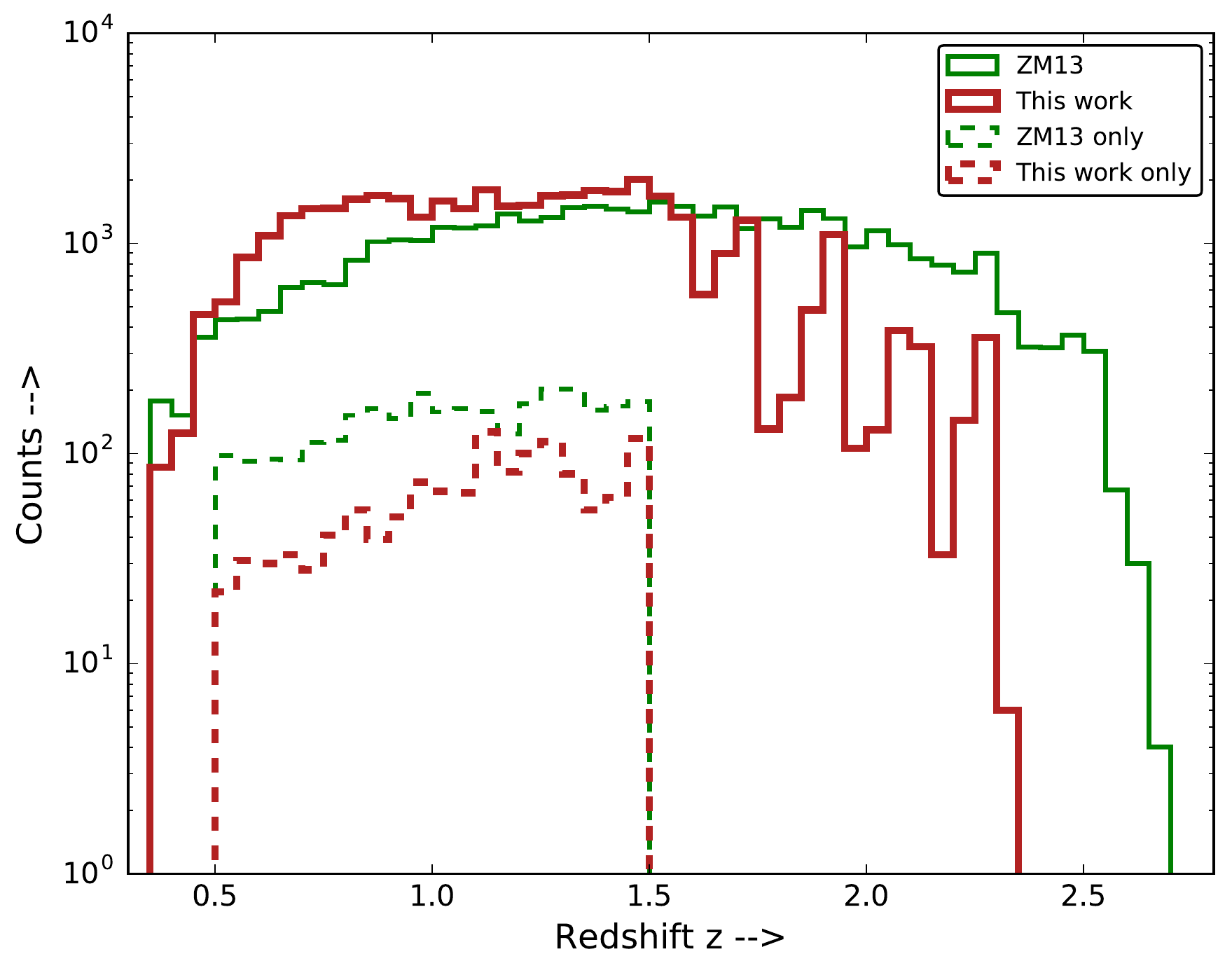}
\caption{The redshift distributions of the absorption systems in the ZM13 (green) and the current work (red) after removing the plausible \mbox{C{\sc\ IV}} systems. The dashed lines show the systems that are unique in each catalogue in the redshift range $0.5 \le z_{2796} \le 1.5$.}
\label{fig_ew_zhu_comp_zdist}
\end{figure}

Further, the systems which are below the threshold SNR level of the current work and that lie outside the redshift range $0.5 > z_{2796} > 1.5$  were also ignored. After all these cuts the ZM13 catalogue contains 11,100 systems for a direct comparison. The distributions of the absorption line redshifts in the two catalogues are shown in Fig. \ref{fig_ew_zhu_comp_zdist}. The green and red solid lines correspond to absorbers in the ZM13 and current work. The dashed lines show the systems that are unique in each catalogue as discussed below.

\begin{figure*}
\centering
\includegraphics[width=0.9\textwidth, height=0.75\textwidth,clip=]{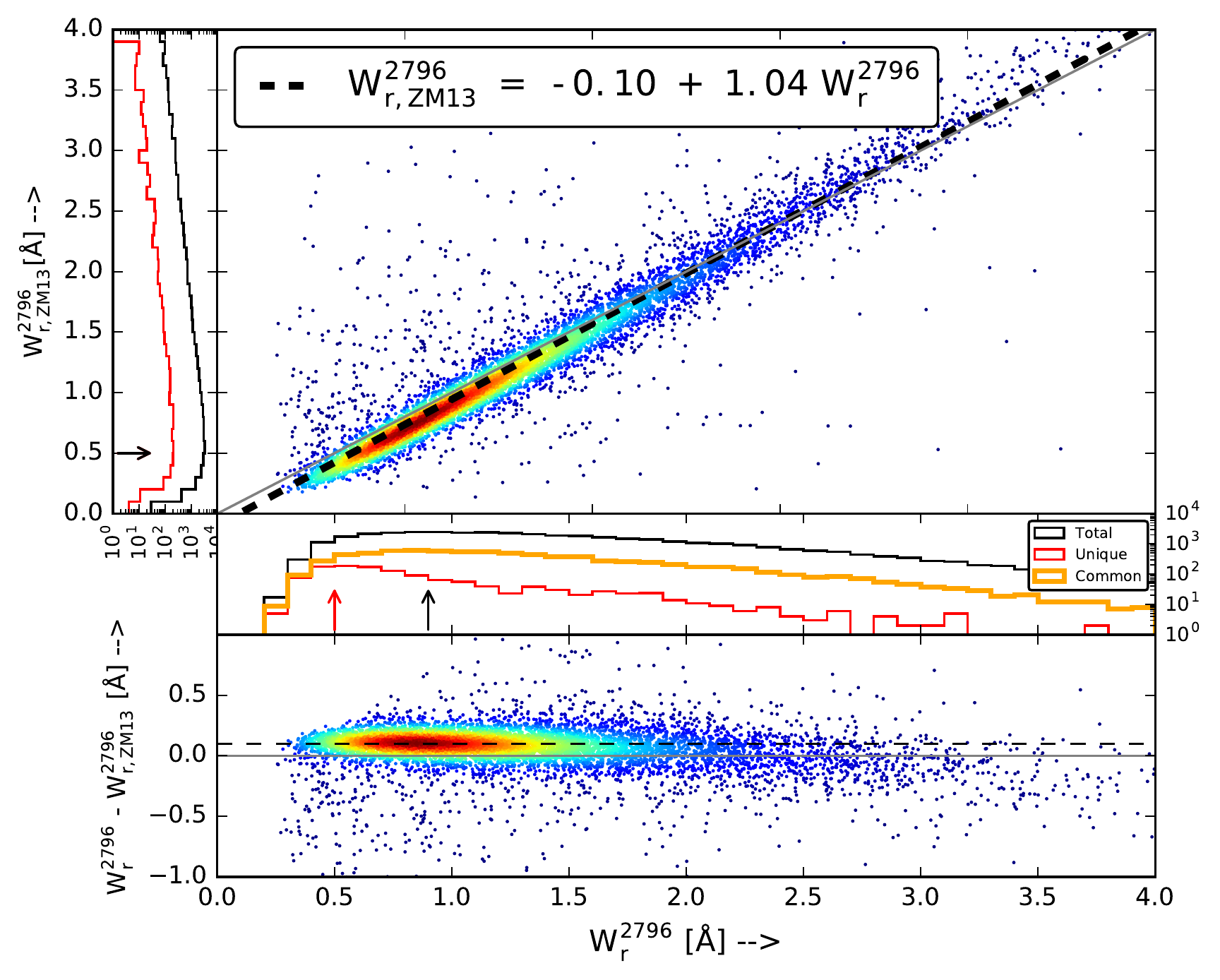}
\caption{The top panel shows the comparison of the EWs between ZM13 and the current work. The black dashed line is the best fit and shows an offset of $0.1\ \AA$ between the measured EW values in the two catalogues. The difference between the two EW values is shown in the bottom panel for clarity. Note that in the current work the EWs are measured using the original spectrum instead of the Gaussian profiles. The discrepancy in the measurements reduces to $0.046\ \AA$ if the EW is measured for Gaussian profiles within $\pm$3 Gaussian widths as measured in ZM13. The (horizontal) histogram in the left and middle panel show the EW distribution of the systems in ZM13 and the current work. The black lines represent the distribution of all the systems while the red is for the systems that are absent in the other catalogue. The EW distribution of the common systems as measured in the current work is shown as the orange histogram.}
\label{fig_ew_zhu_comp}
\end{figure*}

The catalogue produced in this work recovers 74 per cent of the ZM13 systems used for the comparison. The number of common systems goes up by two per cent if only absorption systems with $W_{r} \ge 0.5\ \AA$ were considered for the comparison. Three per cent of the missed detections lie close to either \mbox{C{\sc\ III}]} or \mbox{C{\sc\ IV}} emission lines and were removed because of the cut we applied to remove the false detections due to continuum artefacts. The density plot in the top panel of Fig. \ref{fig_ew_zhu_comp} shows the comparison of the EW of the systems common to both the catalogues. A first order polynomial fit (black dashed line) indicates an offset of $0.1\ \AA$ between the two measurements although this is well with in the error bars of the measured EWs. The discrepancy reduces to $0.046\ \AA$ if the EW is measured within $\pm$3 Gaussian widths using Gaussian profiles as measured in ZM13 instead of the original spectrum. The small offset could be due to the differences in the continuum obtained in the two works. For clarity, the difference between the two measurements is shown in the bottom panel. The horizontal histogram in the left panel of Fig. \ref{fig_ew_zhu_comp} shows the EW distribution of ZM13 catalogue. The black and red lines show total and the systems unique in the ZM13 catalogue. The peak values of the EW distributions are marked using arrows.

To determine the number of unique systems in our catalogue we only used the 10,722 absorption systems that we detected in the spectra of the quasars in common to both the works. ZM13 catalogue contains 75 per cent of our systems. However, the percentage of common systems increase to 88 per cent if we relax the SNR condition of ZM13 from 6 and 3 to their original values of 4 and 2 respectively for the two lines of the doublet. The histogram in the middle panel of Fig. \ref{fig_ew_zhu_comp} shows the EW distributions of the total (unique) absorption systems detected in the current work in black (red) colour. The orange histogram is the EW distribution of the systems common to both the catalogues as measured in this work.

\section{Discussion}
\subsection{Statistical properties of the \mbox{Mg{\sc\ II}} catalogue}
\label{stats_evolution_mgii}

To understand the cosmological evolution of the \mbox{Mg{\sc\ II}} absorbers, it is important to calculate the number of systems detected as a function of both redshift and EW. However, since the sensitivity of the \mbox{Mg{\sc\ II}} search in each bin differs for every quasar spectrum, the number of systems alone may not represent the true evolution. The differences in sensitivity arises mainly because of the differences in:
\begin{itemize} 
\item{The useful redshift range of each spectrum (defined by the $Ly\alpha$ forest region, and the quasar redshift);}
\item{The SNR of each spectrum (changes because of the quasar brightness); and}
\item{The SNR at different regions of the same spectrum (primarily due to incomplete sky subtraction).}
\end{itemize} 

\begin{figure*}
\includegraphics[width=0.9\textwidth, height=0.67\textwidth,clip=]{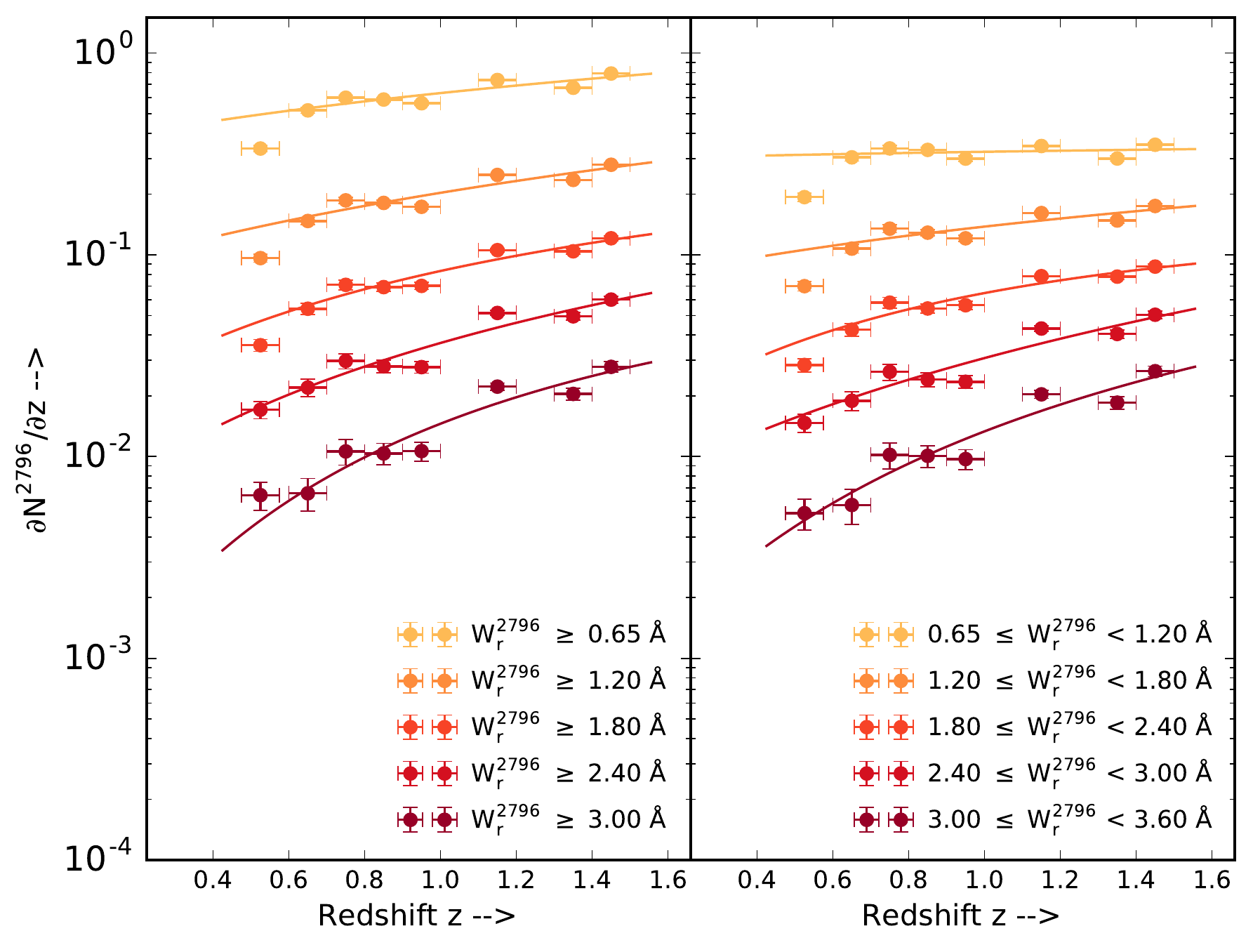}
\caption{Number density evolution $\partial N^{2796}/ \partial z$ of the \mbox{Mg{\sc\ II}} absorption systems for different cumulative (left) and differential (right) EW bins. The solid curves represent the modelling based on Eq. \ref{eq_dn_dz}. The greater evolution of the stronger systems towards low redshifts is evident. The figure is in good agreement with Fig. 10 of ZM13. See the text and Table \ref{dn_dz_table} for more details. }
\label{fig_dn_dz}
\end{figure*}

\begin {table*}
\centering
\caption{Number density $\partial N^{2796}/ \partial z$ values calculated for different EW ranges. The columns represent: EW range; $\partial N^{2796}/ \partial z$ values for the respective EW range at different redshifts; ratio of $\partial N^{2796}/ \partial z$ values at redshifts 0.65 and 1.45.}
\begin {tabular*}{0.8\textwidth}{@{\extracolsep{\fill} } cccccccc}
\hline\\
& & & Redshift z & & & &\\\\
\cline{2-7}\\
EW range & 0.65 & 0.75 & 0.85 & 0.95 & 1.35 & 1.45 & $\partial N^{2796}/ \partial z$\\
& & & & & & & (z$_{0.65}$/z$_{1.45}$)\\
\hline\\
$\partial N^{2796}/ \partial z$ (W$_r\ge$ 0.65\ \AA)  & 0.520  & 0.601  & 0.588  & 0.564  & 0.673  & 0.792  & 0.657 $\pm$ 0.037\\
$\partial N^{2796}/ \partial z$ (W$_r\ge$ 1.20\ \AA)  & 0.147  & 0.186  & 0.181  & 0.173  & 0.235  & 0.280  & 0.526 $\pm$ 0.048\\
$\partial N^{2796}/ \partial z$ (W$_r\ge$ 1.80\ \AA)  & 0.054  & 0.071  & 0.069  & 0.070  & 0.104  & 0.121  & 0.447 $\pm$ 0.070\\
$\partial N^{2796}/ \partial z$ (W$_r\ge$ 2.40\ \AA)  & 0.022  & 0.030  & 0.028  & 0.028  & 0.050  & 0.060  & 0.366 $\pm$ 0.109\\
$\partial N^{2796}/ \partial z$ (W$_r\ge$ 3.00\ \AA)  & 0.007  & 0.011  & 0.010  & 0.011  & 0.020  & 0.028  & 0.236 $\pm$ 0.194\\
\\\hline
\end {tabular*}
\label{dn_dz_table}
\end {table*}

To account for the above effects and derive a homogeneous sample, it is essential to determine the sensitivity of the survey in every redshift and EW bin. The redshift path density $g(W_{j}^{2796},z_{k})$ gives the number of quasar LOS at redshift z$_k$ at which a \mbox{Mg{\sc\ II}} system with EW W$_j$ satisfying all the imposed constraints (see section \ref{sec_cat_ref}) could potentially be detected. As mentioned earlier, we ignored all the possible \mbox{C{\sc\ IV}} systems in the catalogue for the calculations here. Mathematically the redshift path density is represented as 

\begin{eqnarray}
\begin{aligned}
g\left(W_{j}^{2796},z_{k} \right) &=\sum_{i=1}^{N_{QSO}}\ H(z_k - z_{i, min}) \times H(z_{i, max} - z_k )\\
& \times H(W_j^{2796}-W_{i, min}^{2796}(z_{k})) \times H(C_{k,i}-\xi^{2796} \sigma_k))\\
& = \sum_{i=1}^{N_{QSO}}\ H(z_k - z_{i, min})\times H(z_{i, max} - z_k )\\
& \times H(W_j^{2796}-\xi^{2796} {\sigma_{W_{k, i}}}) \times H(C_{k,i} -\xi^{2796} \sigma_{k, i}))\\
\end{aligned}
\label{eq_redshift_path_density_new}
\end{eqnarray} 
where $H$ is the Heaviside step function and the summation is over all the quasar spectra that were searched for the absorption systems. The redshift limits for the \mbox{Mg{\sc\ II}} search in a quasar spectrum $z_{i, min}$ and $z_{i, max}$ were determined by the \mbox{C{\sc\ IV} absorption}\footnote{Since we remove the possible \mbox{C{\sc\ IV}} from the catalogue for estimating the number density evolution, we modify the minimum redshift for the \mbox{Mg{\sc\ II}} search to the location of \mbox{C{\sc\ IV}} absorption instead of the $Ly-\alpha$ for $g(W_{j}^{2796},z_{k})$ calculations.} and \mbox{Mg{\sc\ II}} emission respectively. 

The minimum EW $W_{i, min}(z_{k})$ of the $\lambda$2796 line detected at redshift $z_k$ for a spectrum is given by the detection level $\xi^{2796}=6.0$ times the noise in the measurement $\sigma_{W_k}$ at the desired redshift presented in the rest-frame (see Eq. 2 - 5 of \cite{lanzetta1987}). For the $g\left(W_{j}^{2796},z_{k} \right)$ calculations, we assumed a typical width of ten pixels for the 2796 line. The final term in Eq. \ref{eq_redshift_path_density_new} correspond to ignoring the regions of the respective spectrum with a poor SNR represented using the continuum $C_{k, i}$ and the noise $\sigma_{k, i}$ of the pixels (see section \ref{sec_continuum_noise}) lying in the respective redshift bin.

The total redshift path $g(W_{j}^{2796})$ of the survey for a given EW bin $W_{j}$ is given by integrating the $g\left(W_{j},z_{k} \right)$ over the entire redshift range \citep{lanzetta1987,ellison2004}. For a given redshift bin $\Delta z$ this is
\begin{equation}
g(W_{j}^{2796},z_{1},z_{2})\ =\ \sum_{z_{1}}^{z_{2}} g(W_{j}^{2796},z)\ \Delta z
\label{eq_redshift_path_zbin}
\end{equation}

The $g(W_{j}^{2796},z_1,z_2)$ from Eq. \ref{eq_redshift_path_zbin} can now be used to calculate the number density of the observed absorption systems in the respective redshift bin. 
\begin{equation}
\frac{\partial N^{2796}} {\partial z} (W_{r} \ge W_{j}, z_1, z_2) \ =\ \frac{N (W_r \ge W_j;\ z_1 \le z < z_2)} {g(W_{j}^{2796},z_1,z_2)}
\end{equation} where $N (W_r \ge W_j)$ is the number of systems satisfying the EW threshold and lie in the desired redshift bin. 

Our measurements of the \mbox{$\partial N^{2796}/ \partial z$} for different EW ranges are given in Table \ref{dn_dz_table} and shown in Fig. \ref{fig_dn_dz}. The last column shows the ratio of \mbox{$\partial N^{2796}/ \partial z$} values between the z-range from 0.65 to 1.45. We limited the calculations to $z_{2796}=1.5$ because of the OH band maskings. It is clear that the evolution is steeper for strong \mbox{Mg{\sc\ II}} systems. The number density $\partial N^{2796}/ \partial z$ is an important quantity as it describes the cosmological evolution of the \mbox{Mg{\sc\ II}} absorption systems. We modelled the $\partial N^{2796}/ \partial z$ using \citep{nestor2005}

\begin{equation} 
\frac{\partial^{2}N^{2796}}{\partial z\ \partial W_{r}}=\frac{N(z)}{W(z)}\ exp\left( \frac{W_{r}}{W(z)}\right)
\label{eq_dn_dz}
\end{equation}
where $N(z)=N^* (1+z)^{\alpha}$, and $W(z)=W^* (1+z)^{\beta}$. Note that the EW parameter $W(z)$ also has a power law dependence on the redshift z instead of a simple exponential function. Thus, this is a model for the joint EW-redshift distribution. For more details refer to the original paper by \cite{nestor2005}.

The solid curves in Fig. \ref{fig_dn_dz} are the results from the modelling based on Eq. \ref{eq_dn_dz}. For $W_{r} \ge 1.8\ \AA$ systems, we obtained $N^{*}=1.487$, $W^{*}= 0.31$, $\alpha=0.156$, and $\beta=0.639$ using $\chi^{2}$ minimisation. The steepening of the curves at low redshifts is an indication of the evolution. The evolution is greater for stronger systems with \mbox{W$_r > 1.2 $ \AA} as compared to lower EW systems. In the past, \citet{nestor2005}; \citet{lundgren2009}; \citet{seyffert2013}; and ZM13 have also found a similar evolution of the strong \mbox{Mg{\sc\ II}} systems. In particular, \citet{seyffert2013} and ZM13 combined their observations with the \mbox{Mg{\sc\ II}} systems detected in the near-infrared spectra from a different survey to parameterise the redshift evolution up to redshift z $\sim$ 5. They find a striking correlation between the evolution of the strong \mbox{Mg{\sc\ II}} systems and the star formation history of the universe. Our results (compare Fig. \ref{fig_dn_dz} to Fig. 10 of ZM13) about a steeper evolution of the stronger systems in the local universe agree with all these earlier studies. 

\subsection{Null tests}
\label{sec_null_tests}

\begin{figure*}
\includegraphics[width=0.55\textwidth, height=0.48\textwidth,clip=]{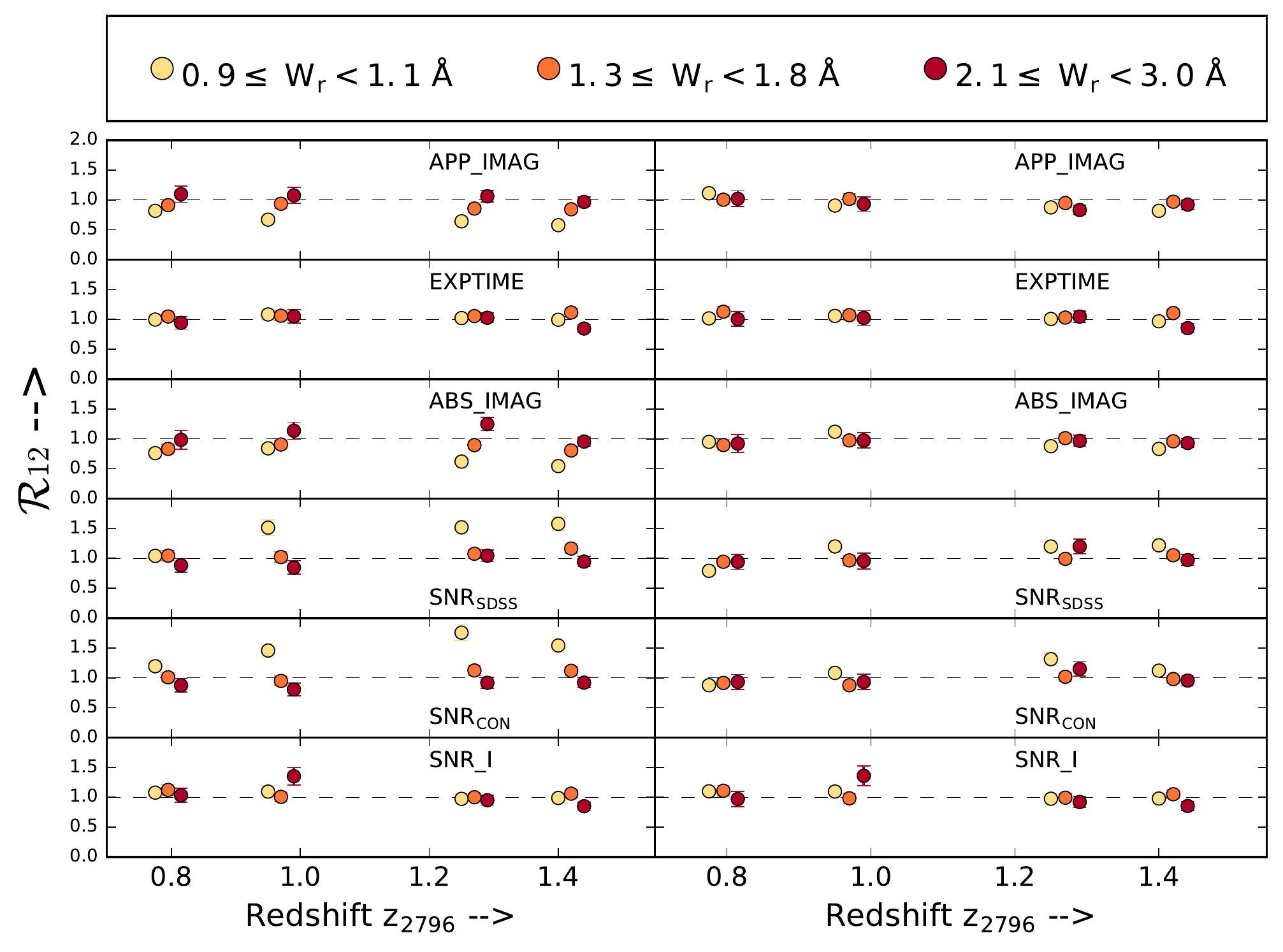}
\includegraphics[width=0.4\textwidth, height=0.4\textwidth,clip=]{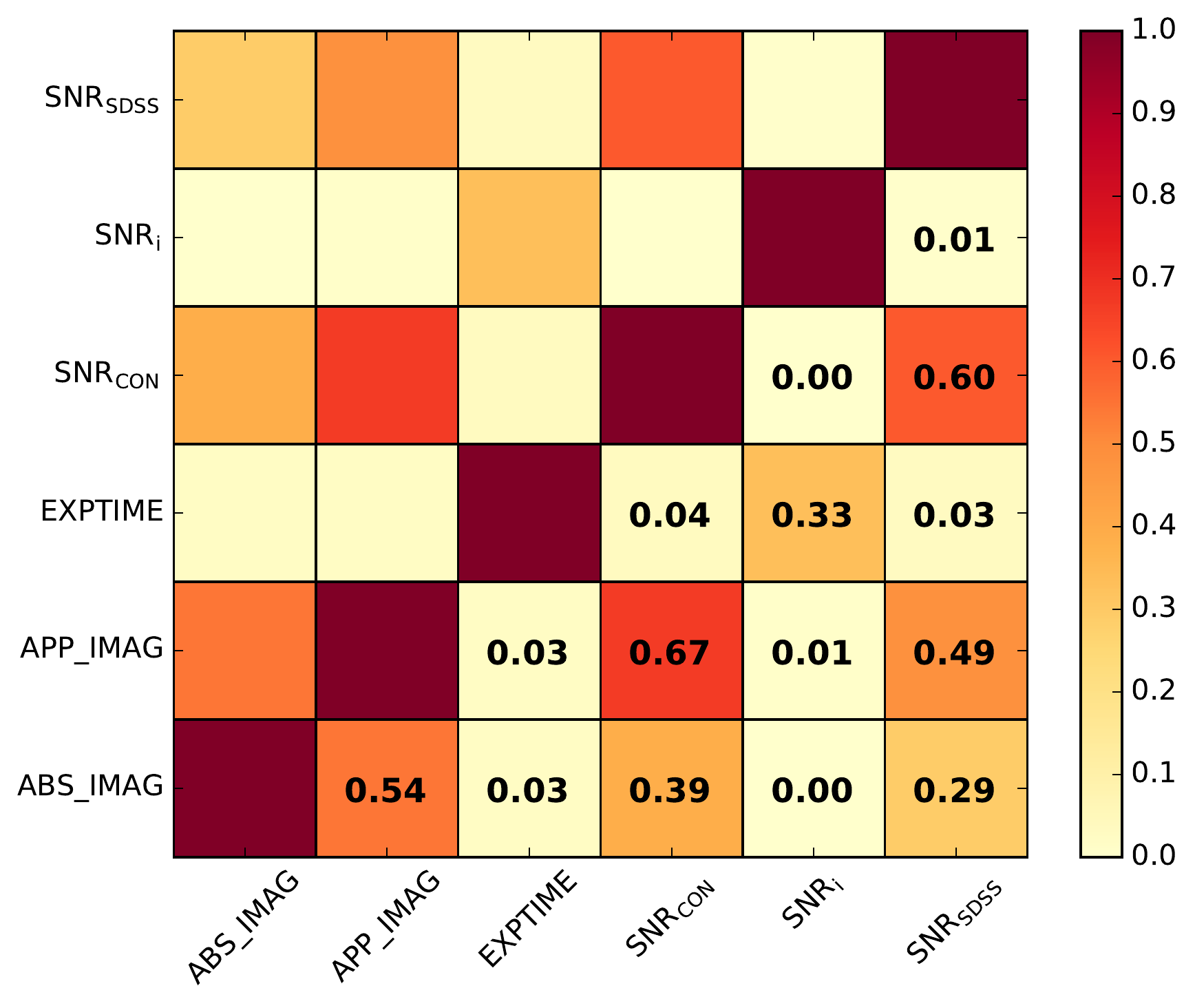}
\caption{(a) The ratio of the number density evolution $\partial N^{2796}/ \partial z$ of the \mbox{Mg{\sc\ II}} absorption systems for different null tests in our null suite before (left) and after (right) the cut based on the $SNR_{CON}$. The large deviation of $\mathcal{R}_{12}$ from unity in the left panel for null tests based on absolute and apparent magnitude, $SNR_{SDSS}$, and $SNR_{CON}$ indicates selection bias. The selection bias is removed (right panel) after removing the quasars with poor spectral SNR. (b) The amount of correlation (Eq. \ref{eq_nt_corr_calc}) between different null tests is shown explicitly.}
\label{fig_null_test}
\end{figure*}

The null tests are statistical tests performed to capture systematic biases in the data. These tests are very common in the CMB research where the raw CMB data is split into two subsets (that differ in the level of contamination due to a particular source of systematic error) and the maps made using the two subsets are differenced to get the final null map. In this work, we determine the ratio of the number density evolution in the two subsets $\mathcal{R}_{12} = \frac{\left(\partial N^{2796}/ \partial z\right)_{1}}{\left(\partial N^{2796}/ \partial z\right)_{2}}$ of the absorption systems for different EW ranges and  types of catalogue splits. If the catalogue is free from systematic bias, then the absolute value of the ratio of the two subsets $\mathcal{R}_{12}$ will be one. A deviation from one will indicate a potential problem that was expected to be captured based on the type of catalogue split (see left panel of Fig. \ref{fig_null_test}). Our null suite consists of six null tests to probe the selection effects due to the spectral quality. They are obtained by splitting the catalogue based on median value of absolute and apparent magnitude of the background quasar, exposure time of the spectrum, average $SNR_{CON}$ of the continuum in the search window as calculated in this work, $SNR_{i,SDSS}$ of the $i$-band, and $SNR_{SDSS}$ of the overall spectrum as derived by the SDSS team. Of course, these data splits are not unique and could be highly correlated. For example, a fainter quasar will surely have a low SNR unless the photons are integrated for a longer duration. In that sense the different null tests do not give distinct information about the systematics. Nevertheless, we show the plots of $\mathcal{R}_{12}$ without including all the null tests for the final $\chi^2$ statistics. In fact, the correlation $C(A,B)$ between two null tests A and B calculated as\footnote{Colin Bischoff, Doctoral dissertation; \url{http://quiet.uchicago.edu/depot/pdf/bischoff\_thesis.pdf}} below and explicitly shown in the right panel of Fig. \ref{fig_null_test}. 

\begin{eqnarray}
\begin{aligned}
C(A,B)&=\sqrt{N(A_1)N(A_2)N(B_1)N(B_2)}\\
& \times\ (T1 - T2 - T3 + T4)
\label{eq_nt_corr_calc}
\end{aligned}
\end{eqnarray}
where
\begin{eqnarray*} 
\begin{aligned}
T1= \frac{N(A_1\cap B_1)} {N(A_1)N(B_1)};\ T2= \frac{N(A_1\cap B_2)} {N(A_1)N(B_2)}\\
T3= \frac{N(A_2\cap B_1)} {N(A_2)N(B_1)};\ T4= \frac{N(A_2\cap B_2)} {N(A_2)N(B_2)}
\end{aligned}
\end{eqnarray*}
and $A_1,B_1,A_2,B_2$ are the first and the second subsets of the two null tests. 

The $\chi_{null}^{2}$ value for all the data points (every EW range in all redshift bins) in all the null tests is

\begin{equation}
\chi_{null}^{2}=\left( \frac{1 - \mathcal{R}_{12}}{\sigma_{\mathcal{R}_{12}}} \right)^{2}
\label{eq_chisq_null}
\end{equation} where the error $\sigma_{\mathcal{R}_{12}}$ is 

\begin{equation}
\sigma_{\mathcal{R}_{12}}=\left|\mathcal{R}_{12} \right| \sqrt{\left(\frac {\sigma_{\partial N^{2796}/\partial z}} {\partial N^{2796}/\partial z} \right)_{1}^{2} + \left(\frac {\sigma_{\partial N^{2796}/\partial z}} {\partial N^{2796}/\partial z} \right)_{2}^{2}}
\label{eq_sigma_r12}
\end{equation}

We chose suitable redshift and differential EW bins for the null tests to avoid the correlation between data points. For each null test we have \emph{4 (z bins) $\times$ 3\ (EW bins) = 12} degrees of freedom. Using the $\chi_{null}^{2}$ value, we calculate the probability to exceed (PTE) of each data point. We did not calculate the combined PTE value of all the null tests because of the high degree of correlation between the null tests. The null test is a success if the PTE values are distributed uniformly between 0 and 1 (consistent with random Gaussian fluctuations). A very high or low $\chi^2$ value will push the PTEs to be distributed either close to 0 or 1.

The ratio plot $\mathcal{R}_{12}$ for different null tests are shown in the Fig. \ref{fig_null_test} (a). The null tests based on the absolute and apparent magnitude, $SNR_{SDSS}$, and $SNR_{CON}$ fail to pass ($\mathcal{R}_{12} \neq 1$) with extremely high $\chi_{null}^{2}$ values. The distribution of the PTEs is shown in the top panel of Fig.\ref{fig_null_test_results} and the failed null tests peak close to zero as expected. The failure indicates a selection bias in the catalogue based on the brightness of the background quasar. The reason and the correction for this failure are discussed in the next section.

\subsection{Null test results}

\begin{figure}
\includegraphics[width=0.45\textwidth, height=0.4\textwidth,clip=]{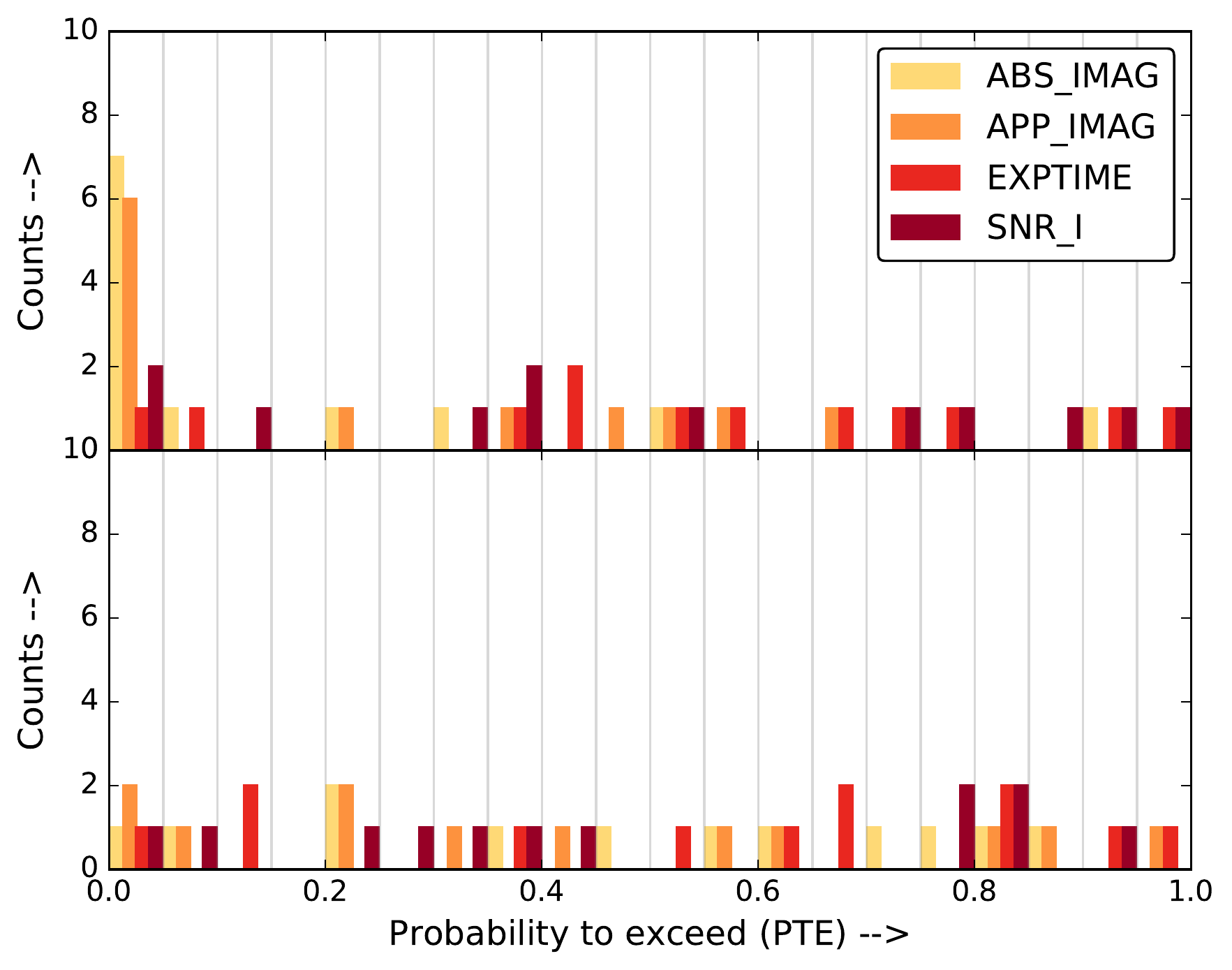}
\caption{The distribution of the PTE values for four null tests before (top) and after (bottom) the $SNR_{CON}$ cut. The peak near PTE value of zero in the top panel is because of (high $\chi^2_{null}$ values) the deviation of the ratio of $\partial N^{2796}/ \partial z$ in the two subsets ($\mathcal{R}_{12}$) from unity. The PTE distribution looks uniform after ignoring the quasar spectra with poor SNR. The indicates that the null tests are successful and the absence of selection bias in the catalogue.}
\label{fig_null_test_results}
\end{figure}

Here we discuss the reasons for the null test failures and check if introducing additional constraints to the catalogue can fix them. In the left panel of Fig. \ref{fig_null_test}, the rather large deviation of the ratio $\mathcal{R}_{12}$ from the expected value of one for the $ABS\_IMAG,\ APP\_IMAG, SNR_{CON}$, and $SNR_{SDSS}$ suggests that the absorber number density depends on the brightness (and hence the spectral SNR) of the background quasar. This is due to the inclusion of the quasars with poor spectral SNR in our analysis. The redshift path $g(W_{j}^{2796})$ is explicitly used to suppress such selection effects and reveal the correct astrophysics of the absorption systems. But the continuum and the noise models for the spectra with poor SNR cannot be highly trusted and could affect the line detection algorithm significantly. This could lead to selection effects in the $\partial N^{2796}/ \partial z$ of the absorption systems detected in the spectra of fainter quasars (as a faint quasar will have a poor SNR compared with a brighter quasar). 

This was indeed the case and the selection bias vanished when we removed quasar spectra with $SNR_{CON} < 5.0$ from the analysis. The distribution of $\mathcal{R}_{12}$ for different null tests also improved and now centered around unity (see right panel of Fig. \ref{fig_null_test} (a)). There was a substantial improvement in the $\chi_{null}^{2}$  values of all the failed null tests after introducing the $SNR_{CON}$ cut. Fig. \ref{fig_null_test_results} shows the distribution of the PTE values before (top panel) and after (bottom panel) the $SNR_{CON}$ cut. The distribution is uniform in the bottom panel hinting the absence of any major selection effects. This cut removes 2803 systems and the catalogue contains 36,981 \mbox{Mg{\sc\ II}} doublets.

In similar studies, \cite{prochter2006b,tejos2009} have reported a higher incidence rate of the strong \mbox{Mg{\sc\ II}} absorbers in the spectra of luminous gamma-ray bursts (GRBs) compared with quasars. \cite{evans2013} have also found a dependence of incidence rate of \mbox{Mg{\sc\ II}} doublets on the quasar luminosity using high resolution quasar spectra. 

Several studies have subsequently attempted to understand this discrepancy (see \cite{cucchiara2009,budzynski2011,wyithe2011} and reference therein) and three main possibilities were considered to explain the phenomenon: (1) the higher incidence rate in GRBs is because of the additional absorbers that are intrinsic to the GRBs, (2) the dusty \mbox{Mg{\sc\ II}} doublets reduce the luminosity of the background quasar such that \mbox{Mg{\sc\ II}} doublet can no longer be detected in the spectra, and (3) the GRBs are lensed by the host galaxy of the \mbox{Mg{\sc\ II}} doublets. 

Using a large sample of GRB spectra, \cite{cucchiara2013} repeated the $\partial N^{2796}/ \partial z$ analysis. Their results agree with the $\partial N^{2796}/ \partial z$ of ZM13 and does not indicate a higher incidence rate of absorbers along the LOS of GRBs. The results from the earlier studies served as a prime motivation for our null tests. Our results are free from selection effects and do not indicate an enhancement of the strong \mbox{Mg{\sc\ II}} absorption lines in the spectra of brighter quasars. 

With the above findings we advise caution when working with the low SNR \mbox{Mg{\sc\ II}} detections in the spectra of faint quasars as the large photometric errors for many candidates selected at the limits of the SDSS imaging survey\footnote{\url{http://www.sdss.org/dr12/algorithms/boss\_quasar\_ts/}} could be detrimental for some projects. Some users may need to perform additional selections depending on their scientific intention.

\section{Conclusion}
We describe the \mbox{Mg{\sc\ II}} absorption catalogue detected using an automated search algorithm from the spectra of the SDSS DR12Q. The detection threshold was SNR $\ge$ 6.0, 3.0 respectively for the two lines of the doublet. The continuum fitting is performed using a mean filter algorithm which was modified with a pseudo-continuum using a median filter to trace the emission lines. The catalogue contains 39,694 systems distributed in the EW range \mbox{0.2 $\le$ W$_{r} \le$ 6.2 \AA} constrained to the redshift range \mbox{$0.35 \le z \le 2.3$}. A separate sky line finder algorithm was employed to remove the strong sky lines in the SDSS spectrum. The catalogue containing the list of sky lines in each quasar spectrum is also publicly available. The SDSS bitmasks were used to eliminate the bad detections in the OH band of the spectrum. Using Gaussian-noise only simulations we estimate $\sim$7.7 per cent of false positives in our catalogue. Our catalogue recovers 76 per cent of the ZM13 absorbers with $W_{r} \ge 0.5\ \AA$. The measurement of the number density $\partial N^{2796}/ \partial z$ of the \mbox{Mg{\sc\ II}} absorbers suggests a steeper evolution of the stronger \mbox{(W$_r \ge$ 1.2 \AA)} \mbox{Mg{\sc\ II}} systems in the low redshift universe as compared with the lower EW systems consistent with other similar works from the earlier data releases of the SDSS. We performed several null tests to quantitatively analyse the dependence of the redshift evolution of the absorption systems on the characteristics of the background quasar. The null test results indicate no selection effects if the quasars with poor spectral $SNR_{CON} <5.0$ are removed from the analysis. The resultant catalogue contains 36,981 systems. The \mbox{Mg{\sc\ II}} absorption catalogue is publicly available and can be downloaded from \url{http://srini.ph.unimelb.edu.au/mgii.php}.

\section{Acknowledgments}
SR acknowledges the CONICYT PhD studentship and the support from CONICYT Anillo project (ACT 1122). SR performed a part of this work at the Aspen Centre for Physics, which is supported by National Science Foundation grant PHY-1066293. SR also acknowledges the support from Australian Research Council's Discovery Projects scheme (DP150103208). The authors thank Isabelle P\^{a}ris for revising the paper draft, and clarifying several questions about the SDSS DR12Q data that were crucial for this work.  SR thanks Prof. Sebastian Lopez for useful discussions about the QAL studies and \mbox{Mg{\sc\ II}} absorbers. The plotting style for some of the plots in this work was inspired from \cite{seyffert2013}; ZM13 for the ease of comparison. The authors thank the anonymous referee for all the useful suggestions which were seminal.

LEC received partial support from the Centre of Excellence in Astrophysics and Associated Technologies (PFB 06), and from a CONICYT Anillo project (ACT 1122). 

This research has used the SDSS DR12Q catalogue \cite{paris2015}. Funding for SDSS-III has been provided by the Alfred P. Sloan Foundation, the Participating Institutions, the National Science Foundation, and the U.S. Department of Energy Office of Science. The SDSS-III web site is http://www.sdss3.org/.

SDSS-III is managed by the Astrophysical Research Consortium for the Participating Institutions of the SDSS-III Collaboration including the University of Arizona, the Brazilian Participation Group, Brookhaven National Laboratory, Carnegie Mellon University, University of Florida, the French Participation Group, the German Participation Group, Harvard University, the Instituto de Astrofisica de Canarias, the Michigan State/Notre Dame/JINA Participation Group, Johns Hopkins University, Lawrence Berkeley National Laboratory, Max Planck Institute for Astrophysics, Max Planck Institute for Extraterrestrial Physics, New Mexico State University, New York University, Ohio State University, Pennsylvania State University, University of Portsmouth, Princeton University, the Spanish Participation Group, University of Tokyo, University of Utah, Vanderbilt University, University of Virginia, University of Washington, and Yale University.






\end{document}